\documentclass[twocolumn,letterpaper,nofootinbib,amsmath]{revtex4}

\pdfoutput=1
\usepackage{epsfig}
\usepackage{subfigure}

\newcommand{\lessthansimilarto}{\lower3pt\hbox{$\buildrel{<}\over{\sim}$}}
\newcommand{\greaterthansimilarto}{\lower3pt\hbox{$\buildrel{>}\over{\sim}$}}

\newcommand{\RR}{\hbox{$I$\kern-3.8pt $R$}}

\begin{document}

\title{Probing the puncture for black hole simulations}

\author{J.~David Brown}
\affiliation{Department of Physics, North Carolina State University,
Raleigh, NC 27695 USA}

\begin{abstract}
With the puncture method for black hole simulations, the second infinity of a wormhole geometry 
is compactified to a single ``puncture point" on the computational grid. The region 
surrounding the puncture quickly evolves to a trumpet geometry. The computational grid covers 
only a portion of the trumpet throat. It ends at a boundary whose location 
depends on resolution.  This raises 
the possibility that perturbations in the trumpet geometry could propagate down the trumpet throat, 
reflect from the puncture boundary, and return to the black hole exterior with a resolution--dependent time delay. 
Such pathological behavior is not observed. This is explained by the observation that 
some perturbative modes propagate in the conformal geometry, others propagate 
in the physical geometry. The puncture boundary exists only in the physical geometry. 
The modes that propagate in the physical geometry are always directed away from the 
computational domain at the puncture boundary. The finite difference 
stencils ensure that these modes are 
advected through the boundary with no coupling to the modes 
that propagate in the conformal geometry. These results are supported by numerical experiments 
with a code that evolves spherically symmetric gravitational fields with standard 
Cartesian finite difference stencils. The code uses  the 
Baumgarte--Shapiro--Shibata--Nakamura formulation of Einstein's equations with 1+log 
slicing and gamma--driver shift conditions. 
\end{abstract}

\maketitle

\section{Introduction}
The puncture method \cite{Campanelli:2005dd,Baker:2005vv} for black hole simulations 
works remarkably well. Initially,
each black hole is represented as a wormhole. The second infinity of each wormhole
is compactified to a point on the computational grid called a puncture. The spatial geometry is evolved 
using finite differencing with the Baumgarte--Shapiro--Shibata--Nakamura (BSSN) 
formulation of Einstein's evolution 
equations \cite{Shibata:1995we,Baumgarte:1998te} and the 
standard gauge conditions consisting of 1+log slicing \cite{Bona:1994dr} and gamma--driver 
shift \cite{Alcubierre:2002kk}.
\begin{figure}[htb] 
	{\includegraphics[scale = 0.85,angle=-90]{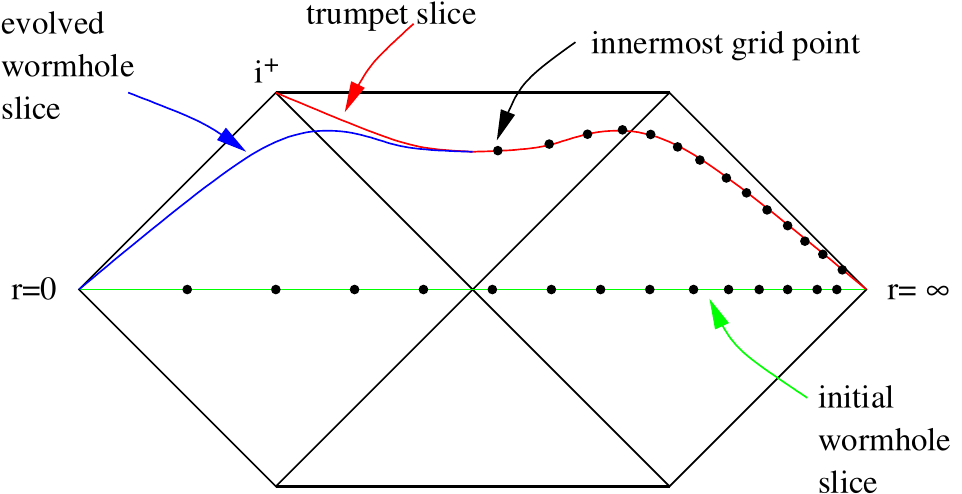}}
	\caption{Penrose diagram showing the relationship between the 
	initial and evolved wormhole slices and a trumpet slice. The heavy dots 
	represent the distribution of numerical grid points. This figure is 
	a sketch based on the results from Refs.~\cite{Brown:2007tb,Brown:2007nt}.}
	\label{penrosefigure}
\end{figure}

The evolution of a single, spherically symmetric puncture black hole is well understood 
from a geometrical point of view \cite{Hannam:2006vv,Brown:2007tb,Brown:2007nt,Hannam:2008sg}. 
Figure \ref{penrosefigure} shows  the Penrose diagram for a Schwarzschild black hole. The initial 
wormhole slice stretches between the two spacelike infinities. The puncture at
$r=0$ coincides with the left spacelike infinity. Heavy dots mark the locations
of points in the computational grid. The puncture itself is not a point in the 
computational grid, since the gravitational field 
diverges there. (Alternatively, one can apply a regularization scheme to keep the 
fields finite at the puncture \cite{Etienne:2007hr,Brown:2007pg,Brown:2008sb}.)

The spatial metric for the initial wormhole slice is written as $ds^2 = \psi^4 (dr^2 + r^2 d\Omega^2)$,
where $d\Omega^2$ is the metric on the unit sphere. The conformal factor $\psi$ diverges like $\psi \sim 1/r$ 
at the puncture. With 1+log slicing and an initially constant lapse, the initial wormhole slice 
evolves to a slice that is left--right symmetric in the Penrose diagram. The gamma--driver shift 
condition, which is built from the conformal metric, breaks the left--right symmetry. 
This allows the  gamma--driver shift to drive the grid points away from the 
left spatial infinity, from left to right in the Penrose diagram. 

The computational grid 
only covers a portion of the evolved wormhole slice. That portion is indistinguishable from 
a portion of a ``trumpet slice" of Schwarzschild. A trumpet slice is a stationary 1+log slice that 
asymptotically approaches the left future infinity $i^+$. 
The conformal factor for a trumpet slice diverges like $\psi \sim 1/\sqrt{r}$. Near 
the puncture $r=0$ the 
trumpet geometry consists of an infinite 
throat with topology $R\times S^2$ and constant cross--sectional area. 

Only a finite portion of the trumpet geometry is covered by the computational grid.  The 
grid ends at the set of grid points that are closest to $r=0$. 
The innermost layer of grid points can be viewed as a discrete {\em boundary} 
that divides the throat into interior and exterior regions. 
As discussed in Ref.~\cite{Brown:2007tb}, the innermost layer of grid points constitute 
a natural excision boundary. The central issue addressed in this 
paper is how we should view this ``puncture boundary". There is, of course, an outer boundary to the 
computational domain 
as well. The numerical relativity community has spent, and continues to spend, a good deal of effort in 
designing boundary conditions for the outer boundary. (See, for example, Ref.~\cite{Sarbach:2007hd}.) 
Should we also worry about boundary conditions 
at the puncture boundary?

One might argue that the puncture boundary cannot affect the exterior evolution of the black hole 
since it lies entirely in the black hole interior. The situation is not that simple. 
We know from the analysis of hyperbolicity for BSSN with standard gauge conditions that 
there are a number of modes that travel with superluminal speeds \cite{Beyer:2004sv,Gundlach:2006tw}. 
These modes represent the gauge freedom in the system, the freedom to change the slicing and 
spatial coordinates. In particular, these modes do not violate the constraints \cite{Brown:2008sb}. Nevertheless, 
it is possible that a gauge wave could
travel into the black hole, reflect from the artificial puncture boundary, and propagate back 
to the black hole exterior. This raises a number of questions. What are the boundary conditions 
at the puncture boundary?  Are they determined by the finite difference stencil? How do 
the boundary conditions affect the reflected waves? 

A more alarming issue is how the grid resolution might affect waves that reflect from the puncture boundary. 
As the computational grid is refined, grid points are added closer to the 
puncture point $r=0$. The innermost layer of grid points is moved farther into the trumpet throat, so 
the distance is increased between the puncture boundary and any finite location, say, 
the black hole horizon. Of course, when we refine the grid  
the coordinate location of the puncture boundary changes by an infinitesimal amount proportional to the 
grid spacing. However, the proper distance from the horizon 
to the puncture boundary changes by a finite amount because the 
geometry diverges at $r=0$. This raises the possibility that wave reflections from the puncture boundary
will be delayed as the grid resolution is increased. 

In Sec.~II I present simulations of a scalar field on a cylinder $R\times S^2$ 
with puncture compactification. Spherically symmetric wave pulses are sent down the cylinder and allowed to reflect
from the puncture boundary.  This system exhibits the pathological behavior described above. Specifically, 
the reflected waves show a clear resolution--dependent time delay. Moreover, we find that the 
form of the reflected wave is affected by the choice of finite difference stencil at the puncture boundary. 

It appears that such pathological behavior does {\em not}  occur 
in black hole simulations that use the puncture method. 
This paper is devoted to explaining why this is so. 
The picture that emerges can be summarized as follows. The initial wormhole data  
evolves very rapidly into a trumpet configuration. For the trumpet geometry, only two of the characteristic fields 
are outgoing at the puncture boundary.\footnote{Throughout this paper I use the terms ``outgoing" and ``outward" to 
mean ``in the positive radial direction; away from the puncture boundary". This differs from the common definition of 
 ``outgoing" as ``from the interior to the exterior of the computational domain".
Likewise,  terms such as ``incoming" and ``inward" will mean ``in the negative radial 
direction; toward the puncture boundary". } These modes have 
sufficiently large positive speeds to allow perturbations to propagate from the puncture boundary 
to the black hole exterior. In effect, these modes
propagate in the {\em conformal geometry}, not the physical geometry. As such, they do not 
sense the movement of the puncture boundary when the grid resolution is changed. The remaining 
characteristic fields  
have negative coordinate speeds at the puncture boundary. If a perturbation occurs in one of these modes, 
the time it takes to reach the boundary will depend on resolution. This raises the possibility that 
one of these incoming modes will be coupled to an outgoing superluminal mode at the puncture boundary and 
give rise to a resolution--dependent reflection.
Numerical simulations indicate that there are no such couplings. This can be understood by considering 
the finite differencing scheme as it appears in both the physical and conformal geometries. 
For the outgoing superluminal modes, the finite difference stencil is equivalent to 
a typical stencil that would be used to evolve smooth fields at the origin of a smooth geometry with topology $R^3$.  
For the modes that propagate in the physical 
geometry, with topology $R\times S^2$, the stencil is one sided at the puncture boundary. 
These ingoing modes are simply advected off the grid with no coupling to the outgoing modes 
or to each other. 

Section III begins with a review of the covariant formulation of BSSN with standard gauge 
conditions \cite{Brown:2009dd}. I then present a detailed derivation of the 
characteristic fields and  
speeds for this system. The analysis uses the ``frozen coefficients" approximation, 
defined by a small amplitude, high frequency limit for perturbations on a background 
solution.  In this way the characteristic fields (or ``modes") are identified as perturbations
that can be realized numerically. I also define a ``gauge system" that has the same superluminal 
characteristics as the full system of BSSN with 1+log slicing and gamma--driver shift.  
The gauge system is linear. 
Useful insights into the full system can be gained by examining the simple gauge system. 

In Sec.~IV I show graphs of the characteristic curves for BSSN with the standard gauge, as 
an initial wormhole geometry evolves to a trumpet. 
Here we see explicitly which modes can carry information from the puncture boundary to the black 
hole exterior. Some of the modes are initially outgoing at the puncture 
boundary, but then quickly change to incoming. In principle, these modes should be fixed by 
boundary conditions while they are outgoing. In practice this does not seem to be a problem, perhaps 
because these modes spend such a short amount of time with their characteristics outgoing at the 
puncture boundary. 

In Sec.~V I describe the results of numerical experiments in which the characteristic fields are 
evolved as perturbations on a single, stationary trumpet geometry. These experiments are used to 
probe the puncture boundary. 
The perturbations consist of a simple wave pulse of compact 
support in one of the incoming modes. 
The system is evolved and reflections in the outgoing modes are 
examined. This technique is used to identify any resolution--dependent time delay 
that might appear in the reflected waves. No such delay is found. 

Section VI contains a summary and discussion of the conclusions that can be drawn from the 
numerical experiments. 

The simulations presented in this paper use a cartoon--type code \cite{Alcubierre:1999ab} 
based on the covariant version of BSSN 
described in Ref.~\cite{Brown:2009dd}. The cartoon code is described in Appendix A. The code is 
designed to evolve spherically symmetric gravitational fields 
using standard Cartesian coordinate finite difference stencils. In Cartesian coordinates, covariant BSSN is completely 
equivalent to the standard BSSN system. With the covariant formulation we can define transformations 
between Cartesian and spherical coordinates in a meaningful way. 

The cartoon code is designed to be third--order convergent. In Appendix A I present the results of a 
two--point convergence test 
with the Hamiltonian constraint and a three--point convergence test with the conformal factor. 
These tests confirm that the code is third--order convergent everywhere in the computational domain. 
It is often stated that puncture evolution codes do not converge at points near the puncture.
Strictly speaking, this claim is incorrect. 
Although there are large finite differencing errors near the puncture,  a properly 
constructed code will be convergent everywhere except at the puncture itself, 
where the conformal factor is not defined.

Appendix B contains a detailed discussion 
of the two techniques that I use to generate the stationary trumpet geometry for the numerical tests. I 
discuss how the numerical data is corrected to account for the fact that neither of these techniques 
yields a geometry that is precisely stationary. 

\section{Puncture evolution of a scalar field}
Consider a massless scalar field $\Phi$ propagating on a three--dimensional cylinder
with topology $R\times S^2$. The metric is $ds^2 = d\zeta^2 + d\theta^2 + \sin^2\theta\, d\phi^2$ where 
$-\infty < \zeta < \infty$ and $\theta$, $\phi$ are spherical coordinates on $S^2$. 
Now compactify the second infinity $\zeta=-\infty$ with the coordinate 
transformation 
\begin{equation}
	\zeta = \sqrt{1 + r^2} + \log\left( \frac{r}{1 + \sqrt{1 + r^2}}\right) \ .
\end{equation}
This transformation maps the three--dimensional cylinder to $R^3$ with a 
puncture at the origin $r=0$. The spatial metric $g_{ab}$ becomes 
\begin{equation}\label{metricforPhi}
	ds^2 = \left( 1 + \frac{1}{r^2}\right) dr^2 +  d\theta^2 + \sin^2\theta\, d\phi^2 \ .
\end{equation}
This metric agrees with the trumpet metric in the 
limit $r\to 0$. In particular, both describe a three--dimensional cylinder with constant cross--sectional area 
and proper length that diverges like $1/r$. 

The scalar field 
equation with unit lapse and vanishing shift is 
\begin{subequations}
\begin{eqnarray}\label{scalarfieldeqns}
	\partial_t \Phi & = & \Pi \ ,\\
	\partial_t \Pi & = & g^{ab} D_a D_b \Phi \ .
\end{eqnarray}
\end{subequations}
Here, $D_a$ is the covariant derivative built from the metric (\ref{metricforPhi}). 

\begin{figure}[htb] 
	{\includegraphics[scale = 0.5]{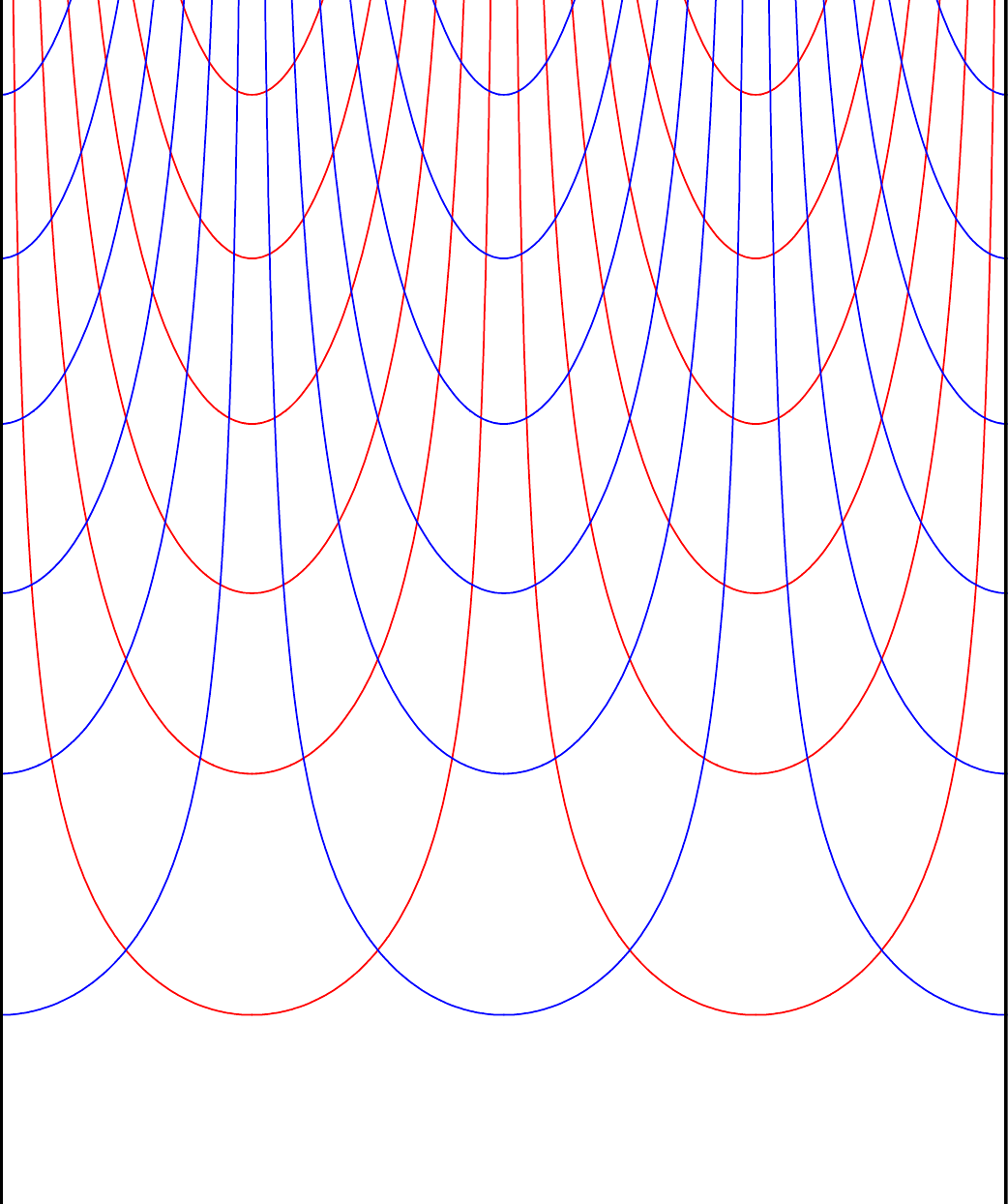}}
	\caption{Coordinates on the compactified cylinder. The left and right edges of this figure are 
	periodically identified. The grid points lie at the intersections of the coordinate lines.}
	\label{CoordsOnCylinder}
\end{figure}

Numerical simulations are carried out using the
Cartesian coordinates defined by $r = \sqrt{x^2 + y^2 + z^2}$, $\cos\theta = z/r$, and $\tan\phi = y/x$. 
The puncture resides at the coordinate origin. 
The numerical grid points lie at the intersections of the coordinate lines
$x/h = 1/2,3/2,\ldots$; $y/h = 1/2,3/2,\ldots$; and $z/h = 1/2,3/2,\ldots$; where $h$ is the 
grid spacing. We can visualize this grid by drawing the coordinate lines 
$x/h = 1/2,3/2,\ldots$ and $y/h = 1/2,3/2,\ldots$ in the equatorial plane defined by $z = 0$. 
Figure \ref{CoordsOnCylinder} shows these coordinate lines plotted on a graph of $\zeta$ versus $\phi$. 
The left and right edges  are identified, so the 
figure represents  an infinitely long two--dimensional cylinder with circumference $2\pi$. 
In this two--dimensional figure, the discrete puncture boundary consists of the 
four points $(x,y) = (\pm h/2, \pm h/2)$ that coincide with the four 
intersections near the bottom of the figure. In the actual three--dimensional simulations, the
puncture boundary consists of the eight grid points $(x,y,z) = (\pm h/2, \pm h/2, \pm h/2)$ 
that are closest to the origin. 

\begin{figure}[htb] 
	{\includegraphics[scale = 1.4,viewport = 80 50 200 180]{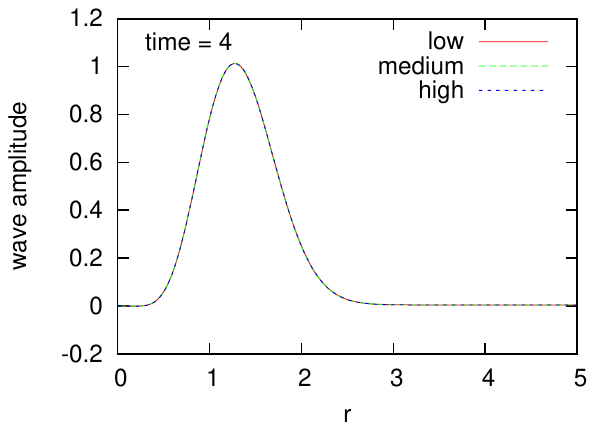}}
	\caption{Scalar field propagating on a punctured cylinder at time $t=4$. Data for
	three resolutions are shown. The three curves overlap at this time.}
	\label{scalarfig1}
\end{figure}

\begin{figure}[htb] 
	{\includegraphics[scale = 1.4,viewport = 80 50 200 180]{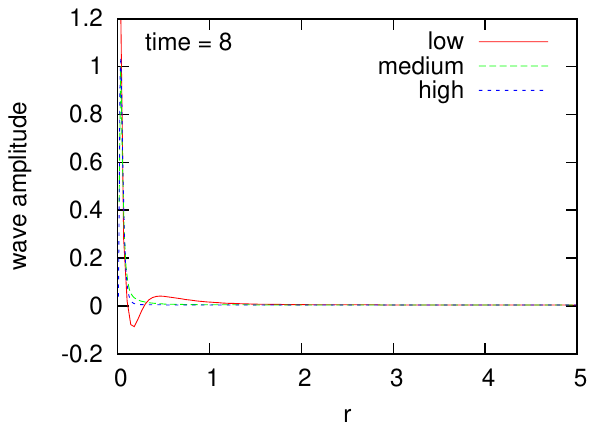}}
	\caption{Scalar field propagating on a punctured cylinder at time $t=8$. 
	The wave is interacting with the puncture boundary.}
	\label{scalarfig2}
\end{figure}

\begin{figure}[htb] 
	{\includegraphics[scale = 1.4,viewport = 80 50 200 180]{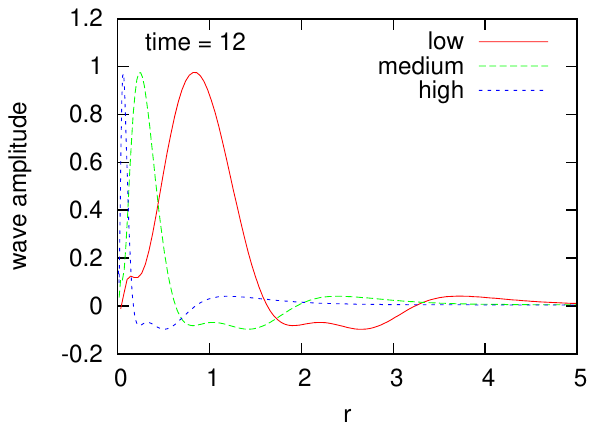}}
	\caption{Scalar field propagating on a punctured cylinder at time $t=12$. 
	The reflected waves are emerging from the puncture.}
	\label{scalarfig3}
\end{figure}

\begin{figure}[htb] 
	{\includegraphics[scale = 1.4,viewport = 80 50 200 180]{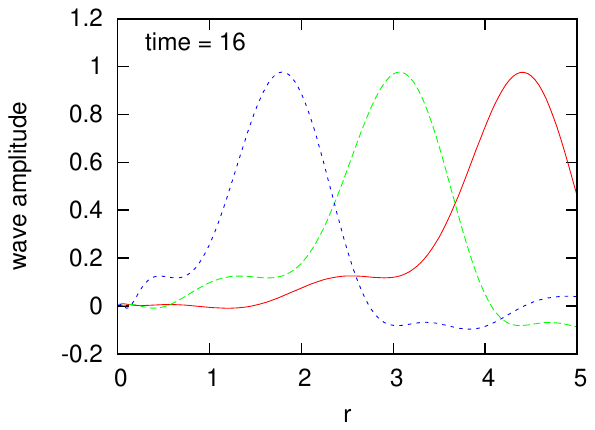}}
	\caption{Scalar field propagating on a punctured cylinder at time $t=16$. 
	The reflected pulse shows a distinct time lag that increases with resolution.}
	\label{scalarfig4}
\end{figure}

Figures \ref{scalarfig1} through \ref{scalarfig4} show a sequence of images taken from simulations at 
three different resolutions, low ($h = 1/25$), 
medium ($h = 1/50$), and high ($h = 1/100$). The initial data consists of a spherical Gaussian
pulse 
\begin{subequations}
\begin{eqnarray}
	\Phi(0) & = & e^{-2(r-5)^2} \ ,\\
	\Pi(0) & = & -4(r-5)e^{-2(r-5)^2} 
\end{eqnarray}
\end{subequations}
that is traveling toward the puncture $r=0$. At time $t=4$, the pulse is still traveling toward the origin 
and the wave forms obtained from the three simulations coincide. The wave pulse hits the puncture boundary
around time $t=8$. The figures at times $t=12$ and $t=16$ show the reflected pulse propagating away from 
the origin. There is a clear dependence on resolution, with the low resolution pulse being reflected most quickly. 
The reflection of the medium resolution pulse is delayed relative to the low resolution case, and the reflection 
of the high resolution pulse is delayed even further.

The resolution dependence of the reflected pulse is precisely what we expect, based on the following analysis.  
Scalar fields  
propagate with proper speed $d\ell/dt = \pm 1$, where $d\ell$ is proper length. Then the coordinate speeds 
in the radial direction are $dr/dt = \pm 1/\sqrt{g_{rr}}$. 
With the metric (\ref{metricforPhi}), the wave speeds are  $v_{in} = - r/\sqrt{1+r^2}$ for incoming waves and 
$v_{out} = r/\sqrt{1+r^2}$ for outgoing waves. 
The time required for a pulse to propagate 
from some large radius $r_0$ to the innermost grid point at $r\approx h/2$ and return to $r_0$ is
\begin{eqnarray}\label{timeequation}
	T_{h} & = & \int_{r_0}^{h/2} \frac{1}{v_{in}} dr  + 
		 \int_{h/2}^{r_0} \frac{1}{v_{out}} dr \nonumber\\
	 & = & {\rm const} - 2\log(h) \ .
\end{eqnarray}
The constant in this expression is independent of grid spacing $h$. 
It follows that the time difference between resolutions $h$ and $h/2$ is 
$T_{h/2} - T_{h} = 2\log 2 \approx 1.37$. We can compare this with our numerical results. 
At time $t=20$, the reflected Gaussian pulses are peaked at 
$r \approx 8.35$, $6.97$, and $5.61$ for low, medium, and high resolutions, 
respectively. The difference between low and medium is $\Delta r \approx 1.38$, whereas the 
difference between medium and high is $\Delta r \approx 1.36$. At these radial 
distances the wave pulses travel with coordinate speeds very close to unity. Thus, the spatial 
separations of $1.38$ and $1.36$ are remarkably close to what we would expect given a time 
lag of $1.37$. 

It is important to keep in mind that the reflection is artificial. The compactification has 
introduced an inner, ``puncture boundary." For the ideal problem of a scalar field propagating 
on an infinite cylinder, the wave pulse would continue to travel down the cylinder with no reflection. 
We can approach this ideal behavior by increasing the resolution. When we increase
the resolution the puncture boundary is pushed farther down the cylinder and the artificial reflection 
is postponed. In principle,  the artificial reflection can be postponed indefinitely. Unfortunately, this is not 
feasible in practice. For example, if we want the reflected pulse 
to be delayed by a time of, say, $20$, then Eq.~(\ref{timeequation}) shows that the required resolution is 
$h \approx e^{-10} \approx 1/22000$. 

Now consider how the finite difference stencil affects the reflected pulse. 
The simulations presented above use standard centered fourth--order stencils. For example, 
for the first and second derivatives of $\Phi$ with respect to the coordinate $z$, we have
\begin{subequations}\label{fourthorderstencils}
\begin{eqnarray}
	(\partial_z\Phi)_k & = & \frac{1}{12h} ( \Phi_{k-2} - 8\Phi_{k-1} 
		\nonumber\\ & & \qquad\quad
		+ 8\Phi_{k+1} - \Phi_{k+2} ) \ ,\\
	(\partial_z^2 \Phi)_k & = & \frac{1}{12h^2} ( -\Phi_{k-2} + 16\Phi_{k-1} - 30\Phi_k 
		\nonumber\\ & & \qquad\quad
		+ 16\Phi_{k+1} - \Phi_{k+2} ) \ ,
\end{eqnarray}
\end{subequations}
where $k$ labels grid points. These finite difference derivatives are fourth--order accurate. 
Alternatively, we can compute the derivatives of $\Phi$ by 
\begin{subequations}\label{alternatestencils}
\begin{eqnarray}
	(\partial_z\Phi)_k & = & \frac{1}{12h} ( - 3\Phi_{k-1} - 10\Phi_{k}
		+ 18\Phi_{k+1} \nonumber\\ & & \qquad\quad
		- 6\Phi_{k+2} + \Phi_{k+3}  ) \ ,\\
	(\partial_z^2 \Phi)_k & = & \frac{1}{12h^2} ( 11\Phi_{k-1} - 20\Phi_k 
		+ 6\Phi_{k+1} \nonumber\\ & & \qquad\quad
		+ 4\Phi_{k+2} - \Phi_{k+3}  ) \ .
\end{eqnarray}
\end{subequations}
This first derivative is fourth--order accurate; the second derivative is third--order accurate. 

\begin{figure}[htb] 
	{\includegraphics[scale = 1,viewport = 140 50 200 220]{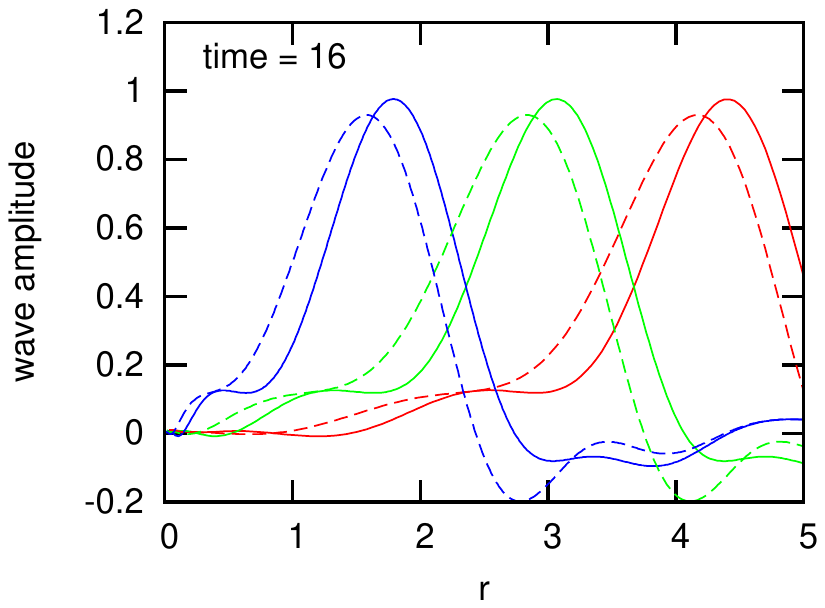}}
	\caption{Reflections from the puncture boundary of a scalar field pulse at three resolutions. 
	The solid curves use the standard centered finite difference stencils. The dashed curves use 
	alternative stencils for the innermost layer of grid points.}
	\label{stencilfig}
\end{figure}

Figure \ref{stencilfig} shows the results of a simulation in which the finite difference stencils are altered for 
the eight grid points closest to the puncture. These are the eight grid points that 
form the discrete puncture boundary. Stencils of the form (\ref{alternatestencils}) were used for 
first and second derivatives along the $x$, $y$, and $z$ coordinate lines. Cross 
derivatives such as $\partial_x\partial_y\Phi$ were not changed.

The solid curves in Fig.~\ref{stencilfig} are a repeat of the curves from Fig.~\ref{scalarfig4}. 
The dashed curves are obtained with the altered stencils applied to the innermost grid points. 
There is a distinct difference between these curves. With the altered stencils, the reflected 
pulses are shifted in time and altered in shape. This suggests that for the puncture evolution 
of a scalar field, the boundary conditions at the puncture boundary are determined, or at 
least affected, by the finite difference stencil.

\section{Perturbation Analysis for BSSN and the standard gauge}
\subsection{Covariant BSSN}
The BSSN equations with 1+log slicing and gamma--driver shift conditions are written in covariant form
in Ref.~\cite{Brown:2009dd},  based on earlier work in Refs.~\cite{Brown:2005aq,Garfinkle:2007yt}. 
The BSSN variables $\varphi$, $g_{ab}$, $K$, and $A_{ab}$ 
are defined in terms of the physical metric $ g_{ab}^{\rm phys}$ and 
extrinsic curvature $ K_{ab}^{\rm phys}$ by 
\begin{subequations}\label{InverseDefinegK}
\begin{eqnarray}
	 g_{ab}^{\rm phys} & = & e^{4\varphi}g_{ab} \ ,\\
	 K_{ab}^{\rm phys} & = & e^{4\varphi}(A_{ab} + g_{ab} K/3) \ .
\end{eqnarray}
\end{subequations}
Let $\Delta \Gamma^a_{bc} \equiv \Gamma^a_{bc} - \mathring\Gamma^a_{bc}$ 
where $\Gamma^a_{bc}$ are the Christoffel symbols built from the conformal metric and 
$\mathring\Gamma^a_{bc}$ are the Christoffel symbols built from a background metric. Also 
use the notation 
$\Delta\Gamma^a \equiv g^{bc} \Delta\Gamma^a_{bc}$. 
The BSSN variables include the ``conformal connection vector" defined by 
\begin{equation}\label{Lambdadef}
	\Lambda^a = \Delta\Gamma^a \ .
\end{equation}
In this paper I will assume that the background is flat. 

The BSSN and standard gauge equations in covariant form are \cite{Brown:2009dd}
\begin{widetext}
\begin{subequations}\label{BSSNeqns}
\begin{eqnarray}
	\partial_t  {g}_{ab}  & = & 2 D_{(a} \beta_{b)} - \frac{2}{3} g_{ab} \bar D_c\beta^c 
		-2{ \alpha} {  A}_{ab}   \ ,\\
    \partial_t {A}_{ab} & = &  \beta^c\partial_c A_{ab} + 2A_{c(a} \partial_{b)}\beta^c 
		- \frac{2}{3} {A}_{ab} \bar D_c\beta^c 
        -2\alpha {A}_{ac}{A}^c_{b} 
       + \alpha {A}_{ab} K   \nonumber\\
       & &  + e^{-4{ \varphi}}  \left[ -2{ \alpha} D_a D_b { \varphi} 
	 + 4{ \alpha} D_a{ \varphi} D_b{ \varphi} + 4 D_{(a}{ \alpha} D_{b)} { \varphi} 
	 - D_a D_b{ \alpha} + { \alpha} {\cal R}_{ab} \right]^{\rm TF} \ ,\\
    \partial_t  { \varphi} 
        & = & \beta^c\partial_c\varphi + \frac{1}{6} \bar D_c\beta^c - \frac{1}{6} {  \alpha} K  \ ,\\
	  \partial_t  K & = & \beta^c\partial_c K +  \frac{\alpha}{3}K^2 + \alpha A_{ab}A^{ab} 
        - e^{-4{ \varphi}} \left( D^2{ \alpha} + 2 D^a{  \alpha} D_a{ \varphi} \right)    \ ,\\
	\partial_t \Lambda^a  & = &  \beta^c \mathring D_c \Lambda^a - \Delta\Gamma^c \mathring D_c\beta^a
	+ g^{bc} {\mathring D}_b{\mathring D}_c \beta^a 
		+\frac{2}{3} \Delta\Gamma^a \bar D_c\beta^c 
        + \frac{1}{3} D^a \bar D_c\beta^c  \nonumber\\ 
	& & -2A^{bc}(\delta_b^a\partial_c \alpha 
		- 6\alpha\delta_b^a\partial_c\varphi - \alpha\Delta\Gamma^a_{bc}) 
	 - \frac{4}{3}\alpha g^{ab} \partial_b K \  ,\\
	\partial_t \alpha & = & \beta^c \mathring D_c \alpha - 2\alpha K \ , \\
	\partial_t\beta^a & = & \beta^c \mathring D_c \beta^a + \frac{3}{4} B^a \ ,\\
	\partial_t B^a & = & \beta^c \mathring D_c B^a 
		+ \left(\partial_t \Lambda^a \right)_{\rm rhs} - \beta^c\mathring D_c \Lambda^a - \eta B^a \ ,
\end{eqnarray}
\end{subequations}
where
\begin{equation}\label{RicciDefinition}
	  {\cal R}_{ab}  \equiv  -\frac{1}{2} g^{cd} \mathring D_c \mathring D_d g_{ab} 
	+ g_{c(a}\mathring D_{b)}\Lambda^c  
	 + g^{de}\Delta\Gamma^c_{de} \Delta\Gamma_{(ab)c} 
	 + g^{cd} \left( 2 \Delta\Gamma^e_{c(a} \Delta\Gamma_{b)ed} 
	+ \Delta\Gamma^e_{ac} \Delta\Gamma_{ebd} \right) \ .
\end{equation}
In the equations above it is assumed that $A \equiv g^{ab}A_{ab} = 0$ is enforced. 
The superscript TF denotes the trace--free part of the expression in 
square brackets, defined with respect to either the conformal metric or the physical metric. 
The term $(\partial_t \Lambda^a)_{\rm rhs}$ in Eq.~(\ref{BSSNeqns}h) must be replaced by the right--hand side of 
Eq.~(\ref{BSSNeqns}e). The operator $D_a$ denotes the covariant derivative 
with respect to the conformal metric $g_{ab}$. The operator $\bar D_a$ denotes the covariant derivative 
with respect to the initial conformal metric, $g_{ab}(0)$. 
The operator $\mathring D_a$ denotes the covariant derivative with respect to the (flat) 
background metric.

If the initial conformal metric is flat and the coordinates are Cartesian, then the 
equations of motion (\ref{BSSNeqns}a--e) above are identical to the original BSSN 
equations \cite{Shibata:1995we,Baumgarte:1998te}.
Likewise, Eq.~(\ref{BSSNeqns}f) is the usual 1+log slicing condition and Eqs.~(\ref{BSSNeqns}g,h)
are the gamma--driver shift conditions. Note that the gamma--driver shift equations 
include all advection terms. With this choice, Eqs.~(\ref{BSSNeqns}g,h) are equivalent 
to the single equation spatial gauge condition used by the Goddard group \cite{vanMeter:2006vi}. 
In the final section I discuss how the results are changed if one or more of the advection 
terms are dropped. 

For the BSSN formulation of Einstein's equations the constraints are defined by 
\begin{subequations}
\begin{eqnarray}
	{\cal H} & = & \frac{2}{3} K^2 - A_{ab} A^{ab} 
	+ e^{-4\varphi}(R - 8D^a\varphi D_a\varphi -8D^2\varphi) \ ,\\
	{\cal M}_a & = &  g^{bc} \mathring D_b A_{ac} - A_{ab} \Delta\Gamma^b - A_c^b \Delta\Gamma^c_{ab}
       + 6A_a^c \partial_c \varphi - \frac{2}{3} \partial_a K  \ ,\\
	{\cal C}^a & = & \Lambda^a - \Delta\Gamma^a \ ,
\end{eqnarray}
\end{subequations}
where $R$ is the conformal Ricci scalar. 
Here, ${\cal H} = 0$ and ${\cal M}_a = 0$ are the usual Hamiltonian and momentum constraints. The 
constraint ${\cal C} = 0$ arises from the definition of the conformal connection vector. 
\end{widetext}

\subsection{Characteristic fields}
Hyperbolicity for systems of partial differential equations with first order time  
and second order 
space derivatives can be a analyzed using the pseudodifferential operator method 
of Refs.~\cite{Kreiss:2001cu,Nagy:2004td}. This technique was first applied to BSSN with 
standard gauge conditions by Beyer and Sarbach \cite{Beyer:2004sv}. 

It will be useful to begin by adopting a condensed notation. 
The BSSN plus standard gauge equations (\ref{BSSNeqns}) have the quasilinear form 
\begin{subequations}\label{quasilinear}
\begin{eqnarray}
	\partial_t q & = & A^a(q) \partial_a q + B(q) v + C(q) \ ,\\
	\partial_t v & = & D^{ab}(q)\partial_a\partial_b q  + E^a(q) \partial_a v + F(q,\partial q,v) \ 
\end{eqnarray}
\end{subequations}
where $q$ represents the ``coordinate variables"
$g_{ab}$, $\varphi$, $\alpha$, and $\beta^a$ and $v$ represents the ``velocity variables" 
$A_{ab}$, $K$, $\Lambda^a$, and $B^a$. The 
coefficients $A^a$, $B$, $C$, $D^{ab}$, and $E^a$ 
depend on the $q$'s. The function $F$ is a quadratic polynomial in the variables 
$\partial_a q$, $v$ with coefficients that depend on the $q$'s. 

Let $\hat q$, $\hat v$ denote a solution of Eqs.~(\ref{quasilinear}) and
consider perturbations $\tilde q$, $\tilde v$ about this solution: 
\begin{subequations}
\begin{eqnarray}
	q & = &  \hat q + \tilde q \ ,\\
	v & = &  \hat v + \tilde v \ .
\end{eqnarray}
\end{subequations}
The  ``frozen coefficients" approximation is defined by a small 
amplitude, high frequency limit as follows. The coefficients in 
Eqs.~(\ref{quasilinear}) and their spatial derivatives are assumed to be of order unity or smaller. 
The perturbation fields 
have small amplitudes proportional to $\epsilon$ and large wave numbers proportional to $k$. 
To be precise, we let $\partial\tilde q \sim \tilde v \sim {\cal O}(\epsilon)$ where $\partial$ 
denotes a space or time derivative. For higher order derivatives, 
$\partial\partial\tilde q \sim \partial\tilde v \sim {\cal O}(\epsilon\cdot k)$. The coordinate perturbation 
$\tilde q$ satisfies  $\tilde q \sim {\cal O}(\epsilon/k)$. 
In the limit as $\epsilon\to 0$ and $k\to\infty$, the leading order (nonvanishing) 
terms in Eqs.~(\ref{quasilinear}) are 
\begin{subequations}\label{principalpart}
\begin{eqnarray}
	\partial_t \tilde q & = &   A^a(\hat q) \partial_a \tilde q + B(\hat q) \tilde v \ ,\\
	\partial_t \tilde v & = & D^{ab}(\hat q)\partial_a\partial_b \tilde q + E^a(\hat q)\partial_a \tilde v \ .
\end{eqnarray}
\end{subequations}
Equation (\ref{principalpart}a) is derived from the ${\cal O}(\epsilon)$ terms in 
Eq.~(\ref{quasilinear}a), and Eq.~(\ref{principalpart}b) is derived from the  
${\cal O}(\epsilon\cdot k)$ terms in Eq.~(\ref{quasilinear}b). 
Equations (\ref{principalpart}) define the principal part of the system (\ref{quasilinear}). 
Note that the coefficients $A^a$, $ B$, $ D^{ab}$, and $ E^a$ are functions of the unperturbed coordinates 
$\hat q$. Below,  I will drop the hats from the unperturbed fields for notational simplicity. 

The hyperbolicity of the system is analyzed by setting the perturbation to a single Fourier mode 
with wave number $k_a$: 
\begin{subequations}\label{fourier}
\begin{eqnarray}
	\tilde q(t,x) = \frac{\breve q}{i|k|} e^{i\omega t + i k_a x^a} \ ,\\
	\tilde v(t,x) = \breve v  e^{i\omega t + i k_a x^a} \ .
\end{eqnarray}
\end{subequations}
Here, $|k| \equiv \sqrt{k_a g^{ab} k_b}$ and $ g_{ab}$ is the unperturbed conformal metric. 
The unit normal to the wave front is denoted $ n_a \equiv k_a/|k|$. The indices on $k_a$ and $n_a$ are  
raised and lowered by the unperturbed conformal metric. 
Note that Eq.~(\ref{fourier}a) implies $n^a\partial_a \tilde q = \breve q e^{i\omega t + i k_a x^a}$. 

Now let $ \mu \equiv \omega/|k|$. The linear Eqs.~(\ref{principalpart}) for the perturbations become
a set of linear algebraic equations for the Fourier coefficients: 
\begin{subequations}
\begin{eqnarray}
	 \mu \breve q & = &  A^a  n_a \,\breve q +  B\, \breve v \ ,\\
	 \mu \breve v & = &  D^{ab} n_a n_b\, \breve q +  E^a n_a \,\breve v \ .
\end{eqnarray}
\end{subequations}
The principal symbol ${\cal P}$ is the matrix defined by the right--hand side of these equations,
\begin{equation}
	{\cal P} = \left(  \begin{array}{cc}  A^a n_a &  B \\  D^{ab} n_a n_b &  E^a n_a 
		\end{array} \right) \ .
\end{equation}
The system of differential equations (\ref{quasilinear}) is strongly hyperbolic if  ${\cal P}$ possesses a 
complete set of eigenvectors with real eigenvalues. 

The characteristic fields for the system are 
obtained from the left eigenvectors of the principal symbol. Let $(\xi,\zeta)$ denote such an eigenvector; 
that is, $(\xi,\zeta){\cal P} =  \mu (\xi,\zeta)$. 
The characteristic field associated with this eigenvector is 
$\chi = \xi n^a\partial_a \tilde q + \zeta\tilde v$. It satisfies 
$\partial_t \chi = \mu n^a\partial_a \chi$ in the frozen coefficients approximation. 

\begin{widetext}
We now spell out the results explicitly for the BSSN plus standard gauge equations. 
The principal parts of these equations are
\begin{subequations}\label{princBSSNeqns}
\begin{eqnarray}
	\partial_t  {\tilde g}_{ab}  & = & \beta^c \partial_c\tilde g_{ab} + 
		2 g_{c(a} \partial_{b)}\tilde\beta^c  - \frac{2}{3} g_{ab} \partial_c\tilde \beta^c 
		-2{ \alpha} { \tilde A}_{ab}   \ ,\\
    \partial_t {\tilde A}_{ab} & = &  \beta^c\partial_c \tilde A_{ab} 
        + e^{-4{ \varphi}}  \left[ -2{ \alpha} \partial_a \partial_b {\tilde \varphi} 
		- \partial_a\partial_b\tilde\alpha 
		- \frac{\alpha}{2} g^{cd}\partial_c\partial_d \tilde g_{ab} 
		+ \alpha g_{c(a}\partial_{b)} \tilde\Lambda^c \right]^{\rm TF} \ ,\\
    \partial_t  {\tilde\varphi} 
        & = & \beta^c\partial_c\tilde \varphi + \frac{1}{6} \partial_c\tilde\beta^c 
	- \frac{1}{6} {  \alpha} \tilde K  \ ,\\
	  \partial_t \tilde K & = & \beta^c\partial_c \tilde K
        - e^{-4{ \varphi}} g^{ab}\partial_a\partial_b { \tilde \alpha}    \ ,\\
	\partial_t \tilde\Lambda^a  & = &  \beta^c \partial_c \tilde\Lambda^a 
	+ g^{bc} \partial_b\partial_c \tilde\beta^a 
        + \frac{1}{3} g^{ab} \partial_b\partial_c\tilde\beta^c 
	 - \frac{4}{3}\alpha g^{ab} \partial_b \tilde K \  ,\\
	\partial_t \tilde\alpha & = & \beta^c \partial_c \tilde \alpha - 2\alpha \tilde K \ , \\
	\partial_t\tilde \beta^a & = & \beta^c \partial_c \tilde\beta^a + \frac{3}{4} \tilde B^a \ ,\\
	\partial_t \tilde B^a & = & \beta^c \partial_c \tilde B^a 
		+ g^{bc} \partial_b\partial_c \tilde\beta^a 
        + \frac{1}{3} g^{ab} \partial_b\partial_c\tilde\beta^c 
	 - \frac{4}{3}\alpha g^{ab} \partial_b \tilde K\ .
\end{eqnarray}
\end{subequations}
Recall that the hats have been dropped from the unperturbed solution. 
As above, we define $|k| \equiv \sqrt{k_a  g^{ab} k_b}$
and $n_a \equiv k_a/|k|$, so 
that $n_a$ is normalized with respect to the unperturbed conformal metric: $n_a  g^{ab} n_b = 1$. 
The principal symbol is defined by 
\begin{subequations}\label{princparts}
\begin{eqnarray}
	\mu  {\breve g}_{ab}  & = & \beta^n \breve g_{ab} + 
		2 n_{(a}  \breve\beta_{b)}  - \frac{2}{3} g_{ab}  \breve \beta^n 
		-2{ \alpha} { \breve A}_{ab}   \ ,\\
    	\mu {\breve A}_{ab} & = &  \beta^n \breve A_{ab} 
        	+ e^{-4{ \varphi}}  \left[ -2{ \alpha} n_a n_b {\breve \varphi} 
		- n_an_b\breve\alpha 
		- \frac{\alpha}{2}  \breve g_{ab} 
		+ \alpha n_{(a} \breve\Lambda_{b)} \right]^{\rm TF} \ ,\\
    	\mu{\breve\varphi} 
        	& = & \beta^n\breve \varphi + \frac{1}{6}\breve\beta^n
		- \frac{1}{6} {  \alpha} \breve K  \ ,\\
	\mu \breve K & = & \beta^n \breve K
        	- e^{-4{ \varphi}}  { \breve \alpha}    \ ,\\
	\mu \breve\Lambda^a  & = &  \beta^n \breve\Lambda^a 
		+    \breve\beta^a 
        	+ \frac{1}{3} n^a  \breve\beta^n
	 	- \frac{4}{3}\alpha n^a \breve K \  ,\\
	\mu \breve\alpha & = & \beta^n \breve \alpha - 2\alpha \breve K \ , \\
	\mu \breve \beta^a & = & \beta^n \breve\beta^a + \frac{3}{4} \breve B^a \ ,\\
	\mu \breve B^a & = & \beta^n \breve B^a 
		+  \breve\beta^a 
        	+ \frac{1}{3} n^a  \breve\beta^n
	 	- \frac{4}{3}\alpha n^a \breve K\ .
\end{eqnarray}
\end{subequations}
Here, the notation $T^{n} \equiv T^{a} n_a$ is used for any tensor $T^a$ contracted with the normal 
covector $n_a$. 

Define  $e^a_A$ by $n_a e^a_A = 0$ and $e^a_A  g_{ab} e^b_B = \delta_{AB}$, 
where the upper case indices $A,B\ldots$ range over the values $1$ and $2$. Thus $e^a_A$ forms 
an orthonormal diad in the subspace orthogonal to $n_a$. Indices on $n_a$ and $e^a_A$ are raised 
and lowered with the 
conformal metric and its inverse. For any tensor $T_a$ we define 
$T_n \equiv T_a n^a= T^a n_a \equiv T^n$ and 
$T_A \equiv T_a e^a_A $. The diad index $A$ is raised and lowered with the 
identity tensor $\delta_{AB}$ and its inverse. 

With this notation we can split the principal Eqs.~(\ref{princparts}) into scalar, vector, and trace--free 
tensor blocks. The scalar block is 
\begin{subequations}
\begin{eqnarray}
	\mu\breve g_{nn} & = & \beta^n \breve g_{nn} + \frac{4}{3}\breve\beta_n 
		- 2\alpha\breve A_{nn} \ ,\\
	\mu\breve g_{AA}  & = & \beta^n\breve g_{AA}  - \frac{4}{3}\breve \beta_n 
		+ 2\alpha \breve A_{nn} \ ,\\
	\mu\breve A_{nn} & = & \beta^n \breve A_{nn} + e^{-4\varphi}\left[ -\frac{4}{3}\alpha\breve\varphi
		- \frac{2}{3}\breve\alpha - \frac{\alpha}{3}\breve g_{nn} + \frac{2}{3}\alpha\breve\Lambda_n
		+ \frac{\alpha}{6}\breve g_{AA} \right] \ ,\\
	\mu\breve\varphi & = & \beta^n \breve\varphi + \frac{1}{6}\breve\beta_n 
		- \frac{1}{6}\alpha\breve K \ ,\\
	\mu\breve K & = & \beta^n \breve K - e^{-4\varphi} \breve\alpha \ ,\\
	\mu\breve\Lambda_n & = & \beta^n \breve\Lambda_n + \frac{4}{3} \breve\beta_n 
		- \frac{4}{3}\alpha \breve K \ ,\\
	\mu\breve\alpha & = & \beta^n \breve\alpha - 2\alpha \breve K \ ,\\
	\mu\breve\beta_n & = & \beta^n \breve\beta_n + \frac{3}{4} \breve B_n \ ,\\
	\mu\breve B_n & = & \beta^n \breve B_n + \frac{4}{3} \breve\beta_n 
		- \frac{4}{3}\alpha \breve K \ ,
\end{eqnarray}
\end{subequations}
where $\breve g_{AA} \equiv \breve g_{AB}\delta^{AB}$. 
In deriving these equations I have used the fact that the perturbation $\tilde A_{ab}$ is trace-free. This implies
$\breve A_{nn} + \breve A_{AA} = 0$ so that $\breve A_{AA}$ can be eliminated in favor of $\breve A_{nn}$. 
The vector block of the principal symbol is 
\begin{subequations}
\begin{eqnarray}
	\mu\breve g_{nA} & = & \beta^n \breve g_{nA} + \breve\beta_A - 2\alpha \breve A_{nA} \ ,\\
	\mu\breve A_{nA} & = &  \beta^n \breve A_{nA} + e^{-4\varphi}\left[  
		-\frac{\alpha}{2} \breve g_{nA} + \frac{\alpha}{2} \breve \Lambda_A \right] \ ,\\
	\mu\breve\Lambda_A & = & \beta^n \breve\Lambda_A + \breve\beta_A  \ ,\\
	\mu\breve\beta_A & = &  \beta^n \breve\beta_A + \frac{3}{4} \breve B_A \ ,\\
	\mu\breve B_A & = &  \beta^n \breve B_A + \breve\beta_A \ .
\end{eqnarray}
\end{subequations}
Finally, the trace--free tensor block is
\begin{subequations}
\begin{eqnarray}
	\mu\breve g_{AB}^{tf} & = &  \beta^n \breve g_{AB}^{tf} - 2\alpha \breve A_{AB}^{tf} \ ,\\
	\mu\breve A_{AB}^{tf} & = &  \beta^n \breve A_{AB}^{tf} - \frac{\alpha}{2} e^{-4\varphi}
		\breve g_{AB}^{tf}  \ ,
\end{eqnarray}
\end{subequations}
where the trace--free part of a tensor $T_{AB}$ is defined by 
$T_{AB}^{tf} \equiv T_{AB} - (T_{CD}\delta^{CD})\delta_{AB}/2$. 

The characteristic fields obtained from the scalar block are
\begin{subequations}\label{scalardefs}
\begin{eqnarray}
	\chi_1 & = & \tilde B^n - 8 \partial_n\tilde\varphi \ ,\\
	\chi_2 & = & \tilde\Lambda^n - 8 \partial_n\tilde\varphi \ ,\\
	\chi_3 & = &  \partial_n \tilde g_{nn}
		+  \partial_n \tilde g_{AA}  \ ,\\
	\chi_{4}^\pm & = & \frac{1}{\sqrt{2\alpha}} e^{-2\varphi} \partial_n\tilde\alpha \pm \tilde K \ ,\\
	\chi_{5}^\pm & = & \mp \frac{3}{2} \tilde A_{nn} \pm \tilde K + e^{-2\varphi} \tilde\Lambda^n 
		+ \frac{1}{4} e^{-2\varphi} \partial_n \tilde g_{AA} - \frac{1}{2} e^{-2\varphi} \partial_n \tilde g_{nn} 
		- 2 e^{-2\varphi} \partial_n \tilde \varphi   \ ,\\
	\chi_{6}^\pm & = & \frac{3}{4} (1 - 2\alpha e^{-4\varphi}) \tilde B^n \pm \alpha\tilde K 
		+ \alpha e^{-4\varphi} \partial_n \tilde\alpha \mp (1 - 2\alpha e^{-4\varphi}) \partial_n\tilde\beta^n
		\ .
\end{eqnarray}
\end{subequations}	
These relations can be inverted for the field perturbations as long as $(1 - 2\alpha e^{-4\varphi}) \ne 0$. 
In that case we have 
\begin{subequations}\label{inversescalar}
\begin{eqnarray}
	\partial_n \tilde g_{nn} & = & -\chi_1 + \frac{4}{3}\chi_2 + \frac{1}{3}\chi_3 
		- \frac{1}{3}(2\alpha)^{2/3} e^{-2\varphi}\frac{(\chi_4^+ + \chi_4^-)}{(1 - 2\alpha e^{-4\varphi})} 
		- \frac{2}{3} e^{2\varphi} (\chi_5^+ + \chi_5^-) 
		+ \frac{2}{3} \frac{(\chi_6^+ + \chi_6^-) }{(1 - 2\alpha e^{-4\varphi})} 
		  \ ,\\
	\partial_n \tilde g_{AA} & = & \chi_1 - \frac{4}{3}\chi_2 + \frac{2}{3}\chi_3 
		+ \frac{1}{3} (2\alpha)^{2/3}e^{-2\varphi}\frac{(\chi_4^+ + \chi_4^-)}{(1 - 2\alpha e^{-4\varphi})} 
		+ \frac{2}{3} e^{2\varphi} (\chi_5^+ + \chi_5^-) 
		- \frac{2}{3} \frac{(\chi_6^+ + \chi_6^-) }{(1 - 2\alpha e^{-4\varphi})} 
		  \ ,\\
	\tilde A_{nn} & = & \frac{1}{3}  (\chi_4^+ - \chi_4^-) 
		+ \frac{1}{3}  (\chi_5^- - \chi_5^+) \ ,\\
	\partial_n \tilde\varphi & = & -\frac{1}{8}\chi_1 
			- \frac{1}{24} (2\alpha)^{3/2} e^{-2\varphi} \frac{(\chi_4^+ + \chi_4^-)}{(1 - 2\alpha e^{-4\varphi})} 
			+ \frac{1}{12} \frac{(\chi_6^+ + \chi_6^-)}{(1 - 2\alpha e^{-4\varphi})}
			 \ ,\\
	\tilde K & = & \frac{1}{2} (\chi_4^+ - \chi_4^-) \ ,\\
	\tilde \Lambda^n & = & -\chi_1 + \chi_2 
		- \frac{1}{3}(2\alpha)^{3/2} e^{-2\varphi} \frac{(\chi_4^+ + \chi_4^-)}{(1 - 2\alpha e^{-4\varphi})} 
		+ \frac{2}{3} \frac{(\chi_6^+ + \chi_6^-)}{(1 - 2\alpha e^{-4\varphi})}
		\ ,\\
	\partial_n\tilde\alpha & = & \frac{\sqrt{2\alpha}}{2} e^{2\varphi} (\chi_4^+ + \chi_4^-) \ ,\\
	\partial_n\tilde\beta^n & = & 
		\frac{\alpha}{2} \frac{(\chi_4^+ - \chi_4^-)}{(1 - 2\alpha e^{-4\varphi})} 
		- \frac{1}{2} \frac{(\chi_6^+ - \chi_6^-)}{(1 - 2\alpha e^{-4\varphi})} 
		 \ ,\\
	\tilde B^n & = & 
		-\frac{1}{3} (2\alpha)^{3/2} e^{-2\varphi} \frac{(\chi_4^+ + \chi_4^-)}{(1 - 2\alpha e^{-4\varphi})} 
		+ \frac{2}{3} \frac{(\chi_6^+ + \chi_6^-)}{(1 - 2\alpha e^{-4\varphi})} \ .
\end{eqnarray}
\end{subequations}
\end{widetext}
The characteristic fields from the vector and trace--free tensor blocks are
\begin{subequations}
\begin{eqnarray}
	\chi_7 & = & \tilde B_A - \tilde\Lambda_A \ ,\\
	\chi_8^\pm & = & \tilde B_A \mp \frac{2}{\sqrt{3}}\partial_n\tilde\beta_A \ ,\\
	\chi_9^\pm & = & \tilde\Lambda_A - \partial_n \tilde g_{nA} \mp 2e^{2\varphi} \tilde A_{nA} \ ,\\
	\chi_{10}^\pm & = & \tilde A_{AB}^{tf} \pm \frac{1}{2} e^{-2\varphi} \partial_n \tilde g_{AB}^{tf} \ .
\end{eqnarray}
\end{subequations}
The inverse relations are 
\begin{subequations}
\begin{eqnarray}
	\partial_n \tilde g_{nA} & = &  -\chi_7 + \frac{1}{2} (\chi_8^+ + \chi_8^- 
		- \chi_9^+ - \chi_9^-)  \ ,\\
	\tilde A_{nA} & = & \frac{1}{4} e^{-2\varphi} (\chi_9^- - \chi_9^+)  \ ,\\
	\tilde\Lambda_A & = &  -\chi_7 + \frac{1}{2}(\chi_8^+ + \chi_8^-) \ ,\\
	\partial_n \tilde\beta_A & = & \frac{\sqrt{3}}{4} (\chi_8^- - \chi_8^+) \ ,\\
	\tilde B_A & = &  \frac{1}{2} (\chi_8^+ + \chi_8^-) \ ,\\
	\partial_n \tilde g_{AB}^{tf} & = & e^{2\varphi}(\chi_{10}^+ - \chi_{10}^-) \ ,\\
	\tilde A_{AB}^{tf} & = & \frac{1}{2} (\chi_{10}^+ + \chi_{10}^-)  \ .
\end{eqnarray}
\end{subequations}
Note that modes $\chi_7$, $\chi_8^\pm$, and $\chi_9^\pm$ are vectors in the subspace 
orthogonal to $n_a$; the index $A$ 
is suppressed above. Likewise, the tensor indices $AB$ and trace--free symbol $tf$ have been 
suppressed on $\chi_{10}^\pm$. 

The perturbations in the lapse and shift (and auxiliary field) are built from 
modes $\chi_4^\pm$ and $\chi_6^\pm$. These can be viewed as gauge modes. They do not violate 
the constraints: In the 
frozen coefficients approximation the perturbations of the constraints are given by 
\begin{subequations}
\begin{eqnarray}
	\tilde{\cal H} & = &  \frac{4}{3} e^{-4\varphi} \partial_n \chi_2 
		- \frac{2}{3}e^{-4\varphi} \partial_n \chi_3  \nonumber\\ & & 
		- \frac{2}{3}  e^{-2\varphi} \partial_n (\chi_5^+ + \chi_5^-) \ ,\\
	\tilde{\cal M}_n & = &  \frac{1}{3} \partial_n (\chi_5^- - \chi_5^+) \ ,\\
	\tilde{\cal M}_A & = & \frac{1}{4} e^{-2\varphi} \partial_n ( \chi_9^- - \chi_9^+) \ ,\\
	\tilde{\cal C}_n & = & -\frac{1}{3}\chi_2 + \frac{1}{6}\chi_3 + \frac{2}{3} e^{2\varphi} (\chi_5^+ + \chi_5^-) \ ,\\
	\tilde{\cal C}_A & = & \frac{1}{2} (\chi_9^+ + \chi_9^-) \ .
\end{eqnarray}
\end{subequations}
These perturbations depend only on modes $\chi_2$, $\chi_3$, $\chi_5^\pm$, $\chi_9^\pm$ and 
their spatial derivatives. 

\subsection{Characteristic speeds}
The phase angle for the Fourier mode (\ref{fourier}) is $\omega t + k_a x^a = |k|(\mu t + n_a x^a)$. 
The surfaces of constant phase satisfy $n_a dx^a/dt = -\mu$. 
Let us choose the wave fronts to coincide with the surfaces $r = {\rm const}$, where $r$ is one of the 
spatial coordinates. Then the covector $k_a$ is proportional to 
$\delta_a^r$ and $n_a = \delta_a^r/\sqrt{g^{rr}}$. The
coordinate speed of the wave is $ dr/dt$, or 
\begin{equation}
	{\hbox{coordinate speed}} = -\mu\sqrt{g^{rr}} \ .
\end{equation} 
This is the wave speed as seen by 
the ``Lagrangian observers" who move along the $\partial/\partial t$ coordinate lines. 

For ``Eulerian observers" who 
are at rest in the spacelike hypersurfaces, the wave speed differs by the addition of the shift. To be 
precise, recall that  
$-\beta^a dt$ is the change during time $dt$ in the spatial coordinate location of an Eulerian observer. 
Then the coordinate distance that the wave front travels in coordinate time $dt$, 
as seen by an Eulerian observer, is the difference: $dr - (-\beta^r dt) = dr + \beta^r dt$.

The proper distance or time between two surfaces $\sigma$ and $\sigma + d\sigma$ is 
$ds = d\sigma/\sqrt{\pm\partial_a\sigma \gamma^{ab} \partial_b\sigma}$, where $\gamma^{ab} $ 
are the contravariant components of a spatial or spacetime metric. This result is derived 
by first constructing the unit normal to the $\sigma = {\rm const}$ surfaces: 
$n^a = \gamma^{ab} \partial_b\sigma / \sqrt{\pm\partial_c\sigma \gamma^{cd} \partial_d\sigma}$. 
The unit normal can be written as $n^a = \partial x^a/\partial s$, where $s$ is proper distance or 
proper time. Now 
compute $\partial\sigma/\partial s = \partial_a\sigma (\partial x^a/\partial s) = 
\sqrt{\pm\partial_a\sigma \gamma^{ab} \partial_b\sigma}$ and the desired result follows. If $\sigma$ is 
one of the coordinates, the proper distance or time becomes $ds = d\sigma/\sqrt{\pm \gamma^{\sigma\sigma}}$. 

The above argument shows that for an Eulerian 
observer, the wave front travels a proper distance $(dr + \beta^r dt)/\sqrt{g^{rr}_{\rm phys}}$ in a proper 
time of $dt/\sqrt{-g^{tt}_{\rm sptm}}$. Here, $g^{tt}_{\rm sptm}$ is the contravariant $tt$ component of the 
spacetime metric, related to the lapse function by $g^{tt}_{\rm sptm} = -1/\alpha^2$. 
Putting this together we find that the proper speed of a given mode is 
$(dr/dt + \beta^r)/(\alpha \sqrt{g^{rr}_{\rm phys}})$, or 
\begin{equation}
	{\hbox{proper speed}} = \frac{e^{2\varphi}}{\alpha\sqrt{g^{rr}}} (\beta^r - \mu\sqrt{g^{rr}} ) \ .
\end{equation}
Note, $r$ can be any one of the spatial coordinates. We will often choose 
$r$ to be the radial coordinate in a spherical coordinate system. 

The speeds of the various modes are listed in Tables I and II. The first line in each table lists the coordinate speed 
of the mode in general. The second line of each table shows the coordinate speed of a radial wave near a trumpet puncture. 
These speeds are obtained from numerical data, which yield 
$\alpha \approx 0.46\,r$, $\beta^r \approx 0.53\,r$, 
$e^{-2\varphi} \approx 0.75\,r$ and $g^{rr} \approx 1$ near $r=0$ for a single, spherically symmetric trumpet with $M=1$. 
(The numerical coefficients should be accurate to within $10\%$.) The third line of each table lists the 
coordinate speed of a radial wave near a wormhole puncture. These speeds are found from the initial data 
for a single, spherically 
symmetric black hole with $M=1$ in isotropic coordinates. They also assume the initial conditions 
$\alpha = 1$ and $\beta^a = 0$ for the lapse and shift. 
Then near a wormhole puncture, we have 
$\alpha \approx 1$, $\beta^r = 0$, $g^{rr} = 1$, and $e^{-2\varphi} \approx 4 r^2$. The fourth 
line of each table lists the proper speed of each mode. The fifth and sixth lines show the  proper speeds 
near trumpet and wormhole punctures. 

\begin{widetext}
The coordinate grid moves with respect to the Eulerian observers. The coordinate speed of
this motion is  $\beta^r$, and the proper speed is $\beta^r/(\alpha\sqrt{g^{rr}_{\rm phys}}) 
= e^{2\varphi}\beta^r/(\alpha \sqrt{g^{rr}})$. Near the puncture boundary of a trumpet slice, 
the proper speed is $\sim 1.5/r$. Near a wormhole puncture boundary the proper speed is zero, 
assuming the shift vector vanishes. 
\begin{table}
\caption{Coordinate and proper speeds for the scalar modes of BSSN with 1+log slicing and gamma--driver shift.}
\begin{ruledtabular}
\begin{tabular}{c|cccccc}
	 & $\chi_1$ & $\chi_2$ & $\chi_3$ & $\chi_{4}^\pm$ & $\chi_{5}^\pm$ 
		& $\chi_{6}^\pm$ \\ \hline
	coordinate speed & $-\beta^r$ & $-\beta^r$ & $-\beta^r$ 
			& $-\beta^r \pm e^{-2\varphi}\sqrt{2\alpha g^{rr}}$ 
			& $-\beta^r \pm \alpha e^{-2\varphi} \sqrt{g^{rr}}$ 
			& $-\beta^r \pm \sqrt{g^{rr}}$ \\ 
	near trumpet puncture& $ - 0.53\,r$ & $  - 0.53\,r$ & $  - 0.53\,r$ 
		& $ -0.53\,r \pm 0.72\,r^{3/2}$ & $  -0.53\,r \pm 0.35\,r^2$ 
		& $ -0.53\,r \pm 1$ \\ 
	 near wormhole puncture& $ 0$ & $ 0$ & $  0$ 
		& $  \pm 5.7\, r^2 $ & $  \pm 4.0\,r^2$ 
		& $  \pm 1$ \\ \hline
	proper speed & $0$ & $0$ & $0$ & $\pm \sqrt{2/\alpha}$ & $\pm 1$ & $\pm e^{2\varphi}/\alpha$ \\ 
	 near trumpet puncture& $0$ & $0$  & $0$  & $ \pm 2.1/\sqrt{r}$ & $\pm 1$ & $ \pm 2.9/r^2 $ \\ 
	 near wormhole puncture & $0$ & $0$  & $0$  & $ \pm 1.4$ & $\pm 1$ & $ \pm 0.25/r^2 $ \\ 
\end{tabular} 
\end{ruledtabular}
\end{table}

\begin{table}
\caption{Coordinate and proper speeds for the vector modes ($\chi_{7}$, $\chi_8^\pm$ and $\chi_9^\pm$) and 
trace--free tensor modes ($\chi_{10}^\pm$) of BSSN with 1+log slicing and gamma--driver shift.}
\begin{ruledtabular}
\begin{tabular}{c|cccc}
	 & $\chi_{7}$ & $\chi_8^\pm$ & $\chi_9^\pm$ & $\chi_{10}^\pm$  \\ \hline
	coordinate speed & $-\beta^r$ 
			& $-\beta^r \pm \sqrt{3g^{rr}}/2$ 
			& $-\beta^r \pm \alpha e^{-2\varphi}\sqrt{g^{rr}}$ 
			& $-\beta^r \pm \alpha e^{-2\varphi}\sqrt{g^{rr}}$  \\ 
	near trumpet puncture & $ - 0.53\,r$ & $-0.53\,r \pm 0.87$ & $-0.53\,r \pm 0.35\,r^2$ 
		& $-0.53\,r \pm 0.35\,r^2$ \\ 
	 near wormhole puncture& $ 0$ & $\pm 0.87$ & $ 4.0\,r^2$ 
		& $4.0\,r^2$  \\ \hline
	proper speed & $0$ & $\pm \sqrt{3}e^{2\varphi}/(2\alpha) $ & $\pm 1 $ & $\pm 1$  \\ 
	 near trumpet puncture & $0$ & $2.5/r^2$  & $\pm 1$  & $ \pm 1$  \\ 
	 near wormhole puncture & $0$ & $0.22/r^2$  & $\pm 1$  & $ \pm 1$  \\ 
\end{tabular} 
\end{ruledtabular}
\end{table}

\end{widetext}

\subsection{Gauge System}
Many of the issues that require close examination reside in the ``gauge sector". We define 
the gauge sector by 
the following system of linear partial differential equations: 
\begin{subequations}\label{GaugeSector}
\begin{eqnarray}
	  \partial_t \tilde K & = & \beta^c\partial_c \tilde K
        - e^{-4{ \varphi}} g^{ab}\mathring D_a\mathring D_b { \tilde \alpha}    \ ,\\
	\partial_t \tilde\alpha & = & \beta^c \partial_c \tilde \alpha - 2\alpha \tilde K \ , \\
	\partial_t\tilde \beta^a & = & \beta^c \partial_c \tilde\beta^a + \frac{3}{4} \tilde B^a \ ,\\
	\partial_t \tilde B^a & = & \beta^c \partial_c \tilde B^a 
		+ g^{bc} \mathring D_b\mathring D_c \tilde\beta^a \nonumber\\
	& & 
        + \frac{1}{3} g^{ab} \mathring D_b\mathring D_c\tilde\beta^c 
	 - \frac{4}{3}\alpha g^{ab} \partial_b \tilde K\ .
\end{eqnarray}
\end{subequations}
Here, the fields $g^{ab}$, $\varphi$, $\alpha$ 
and $\beta^a$ are frozen. The principal part of this system coincides with
Eqs.~(\ref{princBSSNeqns}d,f,g,h). 
The characteristic fields are $\chi_{4}^\pm$ and $\chi_{6}^\pm$. 
The inverse relations are written in Eqs.~(\ref{inversescalar}e,g,h,i). 

This system is simple enough to allow us to compute explicitly the time evolution of the characteristic 
fields, with the following results: 
\begin{subequations} \label{modetimederivatives}
\begin{eqnarray}
	  & & \partial_t \chi_{4}^\pm + (-\beta^r \pm e^{-2\varphi}\sqrt{2\alpha g^{rr}}) \partial_r \chi_{4}^\pm
		\nonumber\\ & & \qquad =  L_{4}^\pm(\chi_4) \ ,\\
	& & \partial_t \chi_{6}^\pm + (-\beta^r \pm \sqrt{ g^{rr}}) \partial_r \chi_{6}^\pm
		\nonumber\\ & & \qquad = L_{6}^\pm(\chi_4,\chi_6) \ .
\end{eqnarray}
\end{subequations}
These are the usual advection--type 
equations plus sources. The sources arise from the fact that the characteristic speeds are not constants. 
The source functions $L_{4}^\pm$ are linear in the characteristic fields $\chi_4^\pm$ with coefficients that 
depend on the fixed fields $g^{ab}$, $\varphi$, $\alpha$, $\beta^a$ and their spatial derivatives.
The source functions $L_{6}^\pm$ are linear in the characteristic fields $\chi_4^\pm$ and $\chi_6^\pm$ with coefficients that 
depend on the fixed fields $g^{ab}$, $\varphi$, $\alpha$, $\beta^a$ and their spatial derivatives.

For the full BSSN plus standard gauge system, the time derivatives of $\chi_4^\pm$ and $\chi_6^\pm$ 
include the same terms displayed in Eqs.~(\ref{modetimederivatives}). They will have extra source terms that come 
from the nonlinearities in the equations. What Eqs.(\ref{modetimederivatives}) reveal is that, already at the level 
of the linear system (\ref{GaugeSector}),
there is coupling between modes $\chi_4^\pm$ and $\chi_6^\pm$. In particular, note that modes 
$\chi_4^\pm$ will tend to excite modes $\chi_6^\pm$. This coupling plays an important role in the analysis of 
Section V. 

\section{Characteristic curves}
Consider a single, spherically symmetric (Schwarzschild) black hole evolved with the puncture method. 
The initial data is a 
wormhole in Cartesian coordinates, so the BSSN variables are 
$\varphi = \ln(1 + 1/(2r))$, $g_{ab} = \delta_{ab}$, $K = 0$, 
$A_{ab} = 0$ and $\Lambda^a = 0$ where $r = \sqrt{x^2 + y^2 + z^2}$. The lapse, shift 
and auxiliary variable are given initial values $\alpha = 1$, $\beta^a = 0$, and $B^a =  0$. 
The black hole mass $M$ is set to unity. 

\begin{figure}[htb] 
	{\includegraphics[scale = 1.4,viewport = 80 50 200 180]{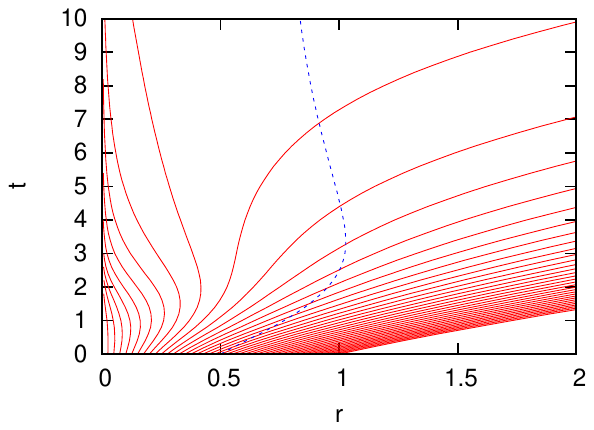}}
	\caption{Characteristic curves for mode $\chi_4^+$. The dashed curve is the black 
	hole horizon.}
	\label{CCfourfigure}
\end{figure}

\begin{figure}[htb] 
	{\includegraphics[scale = 1.4,viewport = 80 50 200 180]{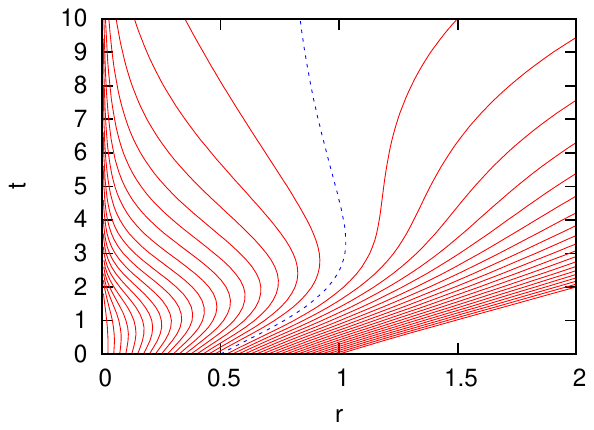}}
	\caption{Characteristic curves for modes $\chi_5^+$, $\chi_9^+$ and $\chi_{10}^+$.}
	\label{CCsixfigure}
\end{figure}

\begin{figure}[htb] 
	{\includegraphics[scale = 1.4,viewport = 80 50 200 180]{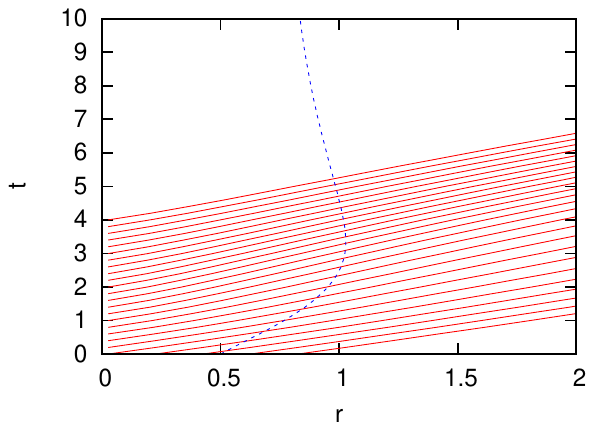}}
	\caption{Characteristic curves for mode $\chi_6^+$.}
	\label{CCeightfigure}
\end{figure}

\begin{figure}[htb] 
	{\includegraphics[scale = 1.4,viewport = 80 50 200 180]{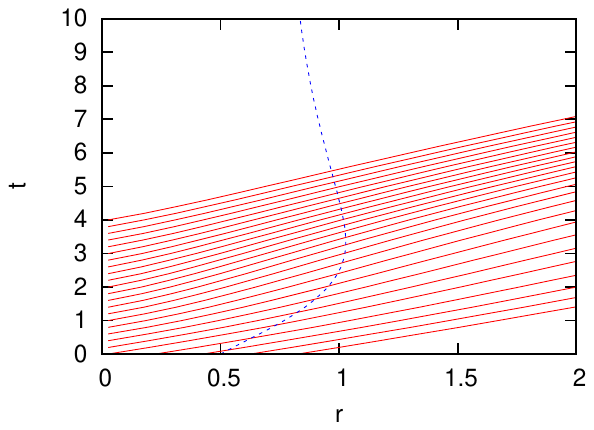}}
	\caption{Characteristic curves for mode $\chi_8^+$.}
	\label{CCelevenfigure}
\end{figure}

Figures \ref{CCfourfigure} through \ref{CCelevenfigure} show the characteristic curves for the various 
modes from time $t=0$ to $t=10$, as the geometry evolves 
from wormhole to trumpet. In each figure the dashed curve is the black hole horizon. 
Figure \ref{CCfourfigure} shows  the characteristic curves for $\chi_4^+$ that begin at time $t=0$ 
at radii $r = 0.025, 0.05, \ldots 1.0$. The modes $\chi_5^+$, $\chi_9^+$, 
and $\chi_{10}^+$ all travel with the speed of light. For these modes Fig.~\ref{CCsixfigure} shows the 
characteristic curves   that 
begin at time $t=0$ and radii $r = 0.025, 0.05, \ldots 1.0$. 
Figures \ref{CCeightfigure} and \ref{CCelevenfigure} show the characteristic curves for 
the outgoing, superluminal modes $\chi_6^+$ and $\chi_8^+$, respectively. In both of these figures, 
one set of curves begin at time $t=0$ and radii $r = 0.025, 0.225, \ldots 0.825$. Another 
set of curves begin at radius $r = 0.025$ and times $t = 0.2, 0.4, \ldots 4.0$. 

It is clear from these figures that perturbations in modes $\chi_6^+$ and $\chi_8^+$ can propagate from the puncture 
boundary to the black hole exterior. Figure \ref{CCfourfigure} shows that mode $\chi_4^+$ is also superluminal; however,  
perturbations within a radius of about $r \approx 0.2$ do not escape to the outside. This result  
was observed in Ref.~\cite{Brown:2008sb}. 

\begin{figure}[htb] 
	{\includegraphics[scale = 1.4,viewport = 80 50 200 180]{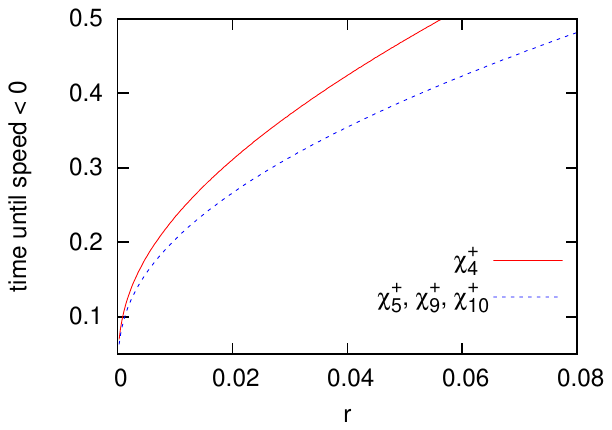}}
	\caption{Time for characteristic speeds to become negative as a function of the radius.}
	\label{timeuntillspeednegative}
\end{figure}

Tables I and II show that modes $\chi_4^+$, $\chi_5^+$, $\chi_9^+$, 
and $\chi_{10}^+$ have positive coordinate speeds near the wormhole puncture. This is difficult to see in 
Figures \ref{CCfourfigure} and \ref{CCsixfigure}. What is clear from these figures is that the 
positive speeds do not last for long. Very quickly, as the geometry shifts from wormhole to trumpet, the speeds 
become negative near the puncture. Figure \ref{timeuntillspeednegative} is a graph of the time required 
for modes $\chi_4^+$, $\chi_5^+$, $\chi_9^+$, 
and $\chi_{10}^+$ to acquire a negative speed. The time is plotted as a function of $r$, which should 
be viewed as the puncture boundary radius. For example, 
with a grid spacing of $h =1/25$ the  puncture boundary has a radius of $r\approx 1/50 = 0.02$. 
Then Fig.~\ref{timeuntillspeednegative} shows that mode $\chi_4^+$ has a positive speed at the puncture 
boundary up to time $t \approx 0.3$. Beyond $t\approx 0.3$, the characteristic speed for $\chi_4^+$ is negative.  

In principle, any characteristic mode that can propagate into the computational domain (in our terminology, 
any mode that is outgoing at the puncture) should be fixed by boundary conditions.  Typically, a numerical 
code that fails to fix such a mode will be unstable. Thus, one would 
expect that boundary conditions are needed for modes $\chi_4^+$, $\chi_5^+$, $\chi_9^+$, 
and $\chi_{10}^+$ during the early stages of their evolution. Yet, the puncture evolution scheme does not
obviously provide any such
boundary conditions. In practice this does not appear to matter. It seems likely that 
these modes have positive speeds for such a short amount of time that instabilities do not have a chance to grow. 
The poor resolution near the puncture boundary might keep the instabilities' growth rates very low. 

Consider again a simulation with $h=1/25$. In this case the modes $\chi_4^+$, $\chi_5^+$, $\chi_9^+$, 
and $\chi_{10}^+$ have positive speeds and are ``unfixed" for a time of $t\approx 0.3$. 
If the Courant factor is, say $1/4$, then the numerical code will execute about $30$ time steps before the
speeds change sign. This is perhaps too few time steps for an instability to grow to a significant level. 

\section{Perturbations of the trumpet geometry}
In this section we carry out simulations of waves reflecting off the puncture boundary of a 
trumpet black hole geometry with $M=1$. The simulations assume spherical symmetry and use the cartoon code described in 
Appendix A. The unperturbed trumpet data is discussed in Appendix B. 
The perturbations in the BSSN and gauge variables are defined by setting one of the incoming modes to 
\begin{equation}
	\chi = -0.0008\, (r-r_0) e^{-12(r-r_0)^2}
\end{equation}
and solving Eqs.~(\ref{inversescalar}). 
Here, $\chi$ is one of $\chi_1$, $\chi_2$, $\chi_3$, $\chi_4^-$, $\chi_5^-$, or $\chi_6^-$. 
Note that the 
vector and trace--free tensor modes $7$ through $10$ do not exist in spherical symmetry. 
Thus, the numerical tests performed here are restricted to modes $1$ through $6$. 

For the incoming modes $\chi_4^-$, $\chi_5^-$ and $\chi_6^-$, we choose $r_0 = 10$. For 
incoming modes $\chi_1$, $\chi_2$ and $\chi_3$ we use $r_0 = 4.0$. In each case we carefully examine 
the reflections in mode $\chi_6^+$. This is the only mode in spherical symmetry that 
can propagate from the puncture boundary to the black hole exterior. 

Let me be specific about the definition of the characteristic fields in spherical symmetry. The definitions (\ref{scalardefs}) 
include terms such as $\partial_n\tilde g_{nn}$, which is shorthand notation for 
$n^a\partial_a (\tilde g_{bc} n^b n^c)$. The normal vector orthogonal to the $r = {\rm const}$ 
coordinate surfaces 
is $n^a = g^{ar}/\sqrt{g^{rr}}$, which simplifies to $\delta^a_r/\sqrt{g_{rr}}$ in spherical 
symmetry. Thus, we have 
\begin{subequations}\label{ntordefs}
\begin{eqnarray}
	\partial_n\tilde g_{nn} & = & \partial_r ( \tilde g_{rr} / g_{rr} )/\sqrt{g_{rr}} \ ,\\
	\partial_n \tilde g_{AA} & = & 2\partial_r (\tilde g_{\theta\theta}/g_{\theta\theta})/\sqrt{g_{rr}} \ ,\\
	\tilde A_{nn} & = & \tilde A_{rr}/g_{rr} \ ,\\
	\partial_n \tilde\varphi & = & (\partial_r \tilde\varphi)/\sqrt{g_{rr}} \ ,\\
	\tilde K & = & \tilde K \ ,\\
	\tilde\Lambda^n & = & \sqrt{g_{rr}} \tilde\Lambda^r \ ,\\
	\partial_n \tilde\alpha & = & (\partial_r \tilde\alpha)/\sqrt{g_{rr}} \ ,\\
	\partial_n\tilde\beta^n & = & \partial_r (\sqrt{g_{rr}}\tilde\beta^r )/\sqrt{g_{rr}} \ ,\\
	\tilde B^n & = & \sqrt{g_{rr}} \tilde B^r \ .
\end{eqnarray}
\end{subequations}
where $\theta$ is the usual polar angle in spherical coordinates. In the cartoon code described in 
Appendix A, the modes (\ref{scalardefs}) are defined by using the frozen coefficients approximation to 
remove factors of the unperturbed fields $g_{rr}$ and $g_{\theta\theta}$ from the derivatives 
in Eqs.~(\ref{ntordefs}). 
For example, in Eqs.~(\ref{scalardefs}) and (\ref{inversescalar}), the term 
$\partial_n\tilde g_{nn}$ is approximated by $(\partial_r \tilde g_{rr})/(g_{rr})^{3/2}$. 

The figures throughout this section use the following convention: The curves (solid or dashed, with various patterns) 
are obtained from simulations at resolution $h = 1/100$. The data points (dots or crosses or other symbols) 
are obtained from simulations at resolution $h = 1/50$. Unless otherwise stated, only every fifth data point is
displayed for the lower resolution case. 

\subsection{Incident mode $\chi_6^-$}
\begin{figure}[htb] 
	{\includegraphics[scale = 1.4,viewport = 80 50 200 180]{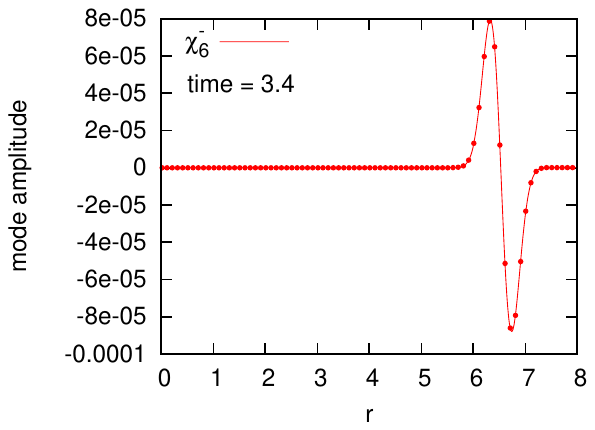}}
	\caption{Incoming mode $\chi_6^-$ at time $3.4$. (The initial excitation is  in 
	 $\chi_6^-$.)}
	 \label{Dmode9chi9at17}
\end{figure}

\begin{figure}[htb] 
	{\includegraphics[scale = 1.4,viewport = 80 50 200 180]{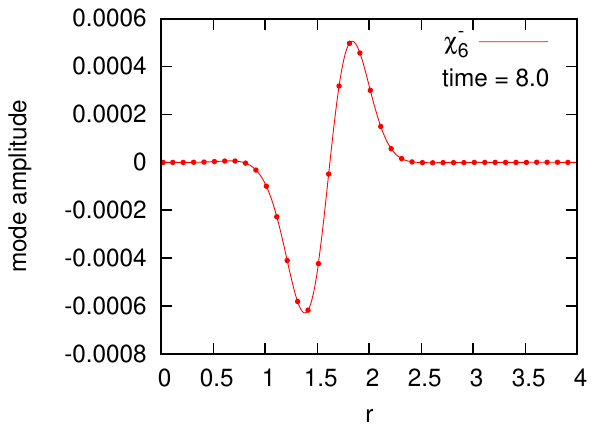}}
	\caption{Incoming mode $\chi_6^-$ at time $8.0$. (The initial excitation is in 
	 $\chi_6^-$.)}
	 \label{Dmode9chi9at40}
\end{figure}

\begin{figure}[htb] 
	{\includegraphics[scale = 1.4,viewport = 80 50 200 180]{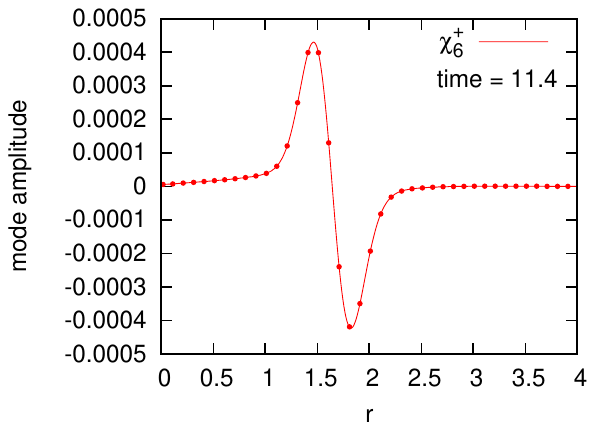}}
	\caption{Outgoing mode $\chi_6^+$ at time $11.4$. (The initial excitation is in 
	 $\chi_6^-$.)}
	 \label{Dmode9chi8at57}
\end{figure}

\begin{figure}[htb] 
	{\includegraphics[scale = 1.4,viewport = 80 50 200 180]{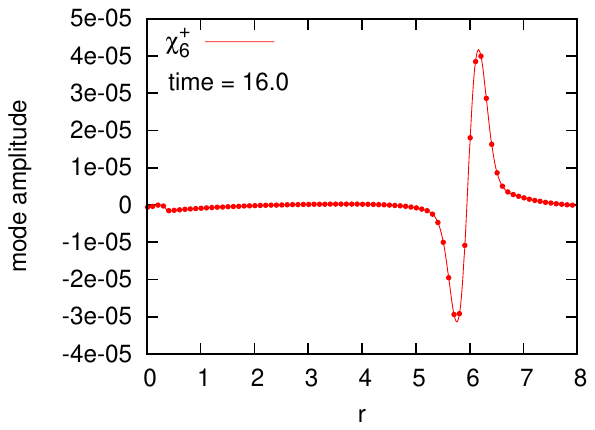}}
	\caption{Outgoing mode $\chi_6^+$ at time $16.0$. (The initial excitation is in 
	 $\chi_6^-$.)}
	 \label{Dmode9chi8at80}
\end{figure}

Figure \ref{Dmode9chi9at17} shows a perturbation in mode $\chi_6^-$ at time $t = 3.4$. 
This pulse propagates inward toward the 
puncture $r=0$. Figure \ref{Dmode9chi9at40} shows the incident pulse at $t = 8.0$, just before it hits the puncture 
boundary. Figures \ref{Dmode9chi8at57} and \ref{Dmode9chi8at80} show the reflected pulse 
$\chi_6^+$ propagating outward at times $11.4$ and $16.0$, respectively.  
In each of these four figures, the dots (every fifth data point from a simulation with $h = 1/50$) lie on top of 
the solid curve (from a simulation with $h=1/100$). These results are convergent. In particular,
the reflected wave pulse does {\em not} show any resolution--dependent time delay. 

It is informative to compare the amplitudes of the incoming and reflected pulses. From Fig~\ref{Dmode9chi9at17} we 
see that the incident mode $\chi_6^-$ has amplitude $A\sim 8\times 10^{-5}$ at $t = 3.4$. By $t = 8.0$ the amplitude has 
grown by almost an order of magnitude to $A\sim 6\times 10^{-4}$. It continues to grow as the pulse approaches the puncture. 
The reflected mode $\chi_6^+$ initially has a very large amplitude. It drops rapidly to $A \sim 4\times 10^{-4}$
at time $t=11.4$, as shown in Fig.~\ref{Dmode9chi8at57}. By $t = 16.0$ it has dropped an order of 
magnitude to $A\sim 4\times 10^{-5}$. At 
equal distances from the puncture (say, $r \approx 6.0$), the reflected pulse is almost as large as the incident 
pulse. 

\begin{figure}[htb] 
	{\includegraphics[scale = 1.4,viewport = 80 50 200 180]{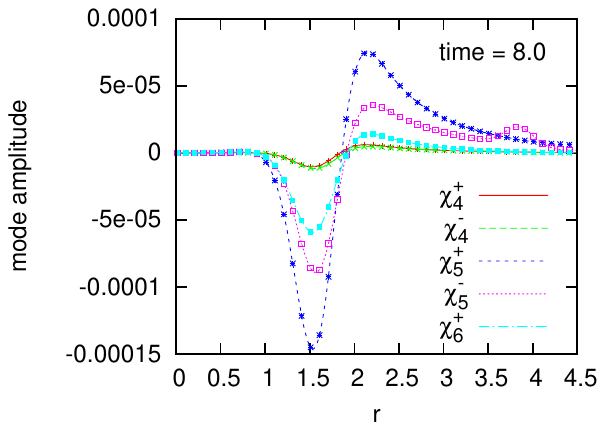}}
	\caption{Modes $\chi_4^\pm$, $\chi_5^\pm$, and $\chi_6^+$ at time $8.0$, 
	carried inward along with the initially excited mode $\chi_6^-$.  Compare with Fig.~\ref{Dmode9chi9at40}.}
	\label{Dmode9chiallat40}
\end{figure}

The small step in the data for $\chi_6^+$ seen in Fig~\ref{Dmode9chi8at80} 
at $r \approx 0.4$ deserves 
some discussion. To begin, let us consider the modes $\chi_4^\pm$, $\chi_5^\pm$ and $\chi_6^+$ before 
the pulse $\chi_6^-$ hits the puncture boundary. Figure \ref{Dmode9chiallat40} displays these modes 
at time $t= 8.0$. If our system of partial differential equations 
were linear with constant wave speeds, these modes would not be excited at all. 
However, due to the nonlinear nature of Einstein's theory, these modes build in amplitude as 
the initial pulse $\chi_6^-$ propagates inward. Even the modes $\chi_4^+$, $\chi_5^+$ and $\chi_6^+$, 
which are normally outgoing, develop  nontrivial profiles that are carried inward along with $\chi_6^-$. 
The largest of these modes is $\chi_5^+$, with an amplitude $A\sim 1\times 10^{-4}$. 
However, keep in mind that relative amplitudes 
among the different modes have little meaning since the normalization of the modes is arbitrary.  
\begin{figure}[htb] 
	{\includegraphics[scale = 1.4,viewport = 80 50 200 180]{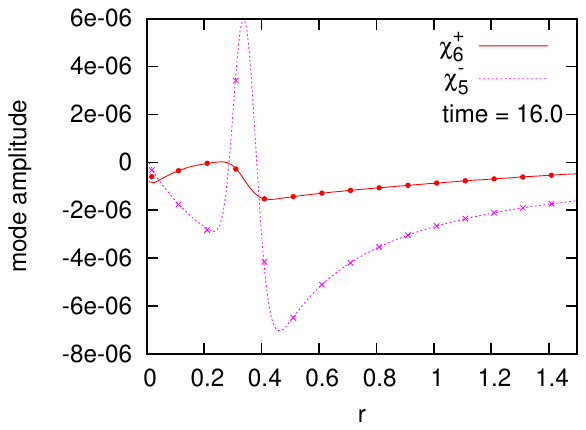}}
	\caption{Modes $\chi_5^-$, and $\chi_6^+$ at time $16.0$. These waveforms 
	are traveling inward. (The initial excitation is in 
	 $\chi_6^-$.)}
	\label{Dmode9chiallat80}
\end{figure}

The important feature of Fig.~\ref{Dmode9chiallat40} is the small hump in $\chi_5^-$ that peaks at about $r = 3.8$. 
This hump is initially part of the larger perturbation in $\chi_5^-$ that is carried along with 
$\chi_6^-$. However, mode $\chi_5^-$ naturally tends to move more slowly than $\chi_6^-$, as seen from the data in  
Table I. By $t=8.0$ a  small piece  of $\chi_5^-$ has broken free from the other perturbations and fallen behind.
It continues to propagate inward, moving more and more slowly as it approaches the puncture. 
Figure \ref{Dmode9chiallat80} shows the hump in $\chi_5^-$ at $t = 16.0$, just before it reaches the puncture. 
That figure includes a plot of $\chi_6^+$. We see that the nonlinear couplings have created a new perturbation in $\chi_6^+$, 
that is carried inward along with $\chi_5^-$. This perturbation in $\chi_6^+$ is seen as a small step at $r \approx 0.4$ 
in both Figs.~\ref{Dmode9chi8at80} and \ref{Dmode9chiallat80}. Note that these features are all convergent. 

\begin{figure}[htb] 
	{\includegraphics[scale = 1.4,viewport = 80 50 200 180]{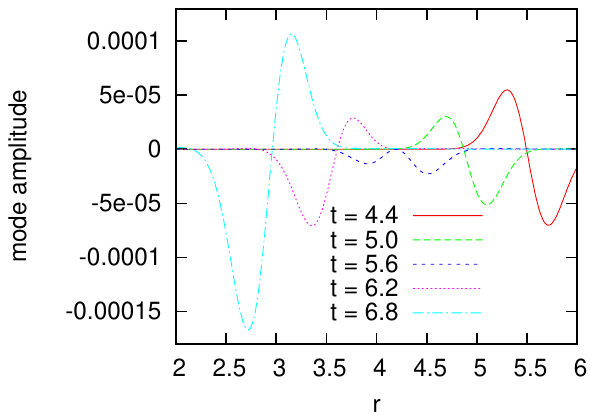}}
	\caption{Time sequence showing mode $\chi_6^-$ as it travels inward through 
	the point $r = 4.06$ where hyperbolicity breaks down.}
	\label{Dinversionfig}
\end{figure}

Finally, let us compare the incoming pulses $\chi_6^-$ at $t = 3.4$ and $t = 8.0$; these are pictured in 
Figs.~\ref{Dmode9chi9at17} and \ref{Dmode9chi9at40}. Observe that the pulse at $t=8.0$ is inverted relative to the 
pulse at $t=3.4$. This is related to the breakdown in hyperbolicity that occurs when 
$(1 - 2\alpha e^{-4\varphi}) = 0$. In particular, the factor 
$(1 - 2\alpha e^{-4\varphi})$ appears in Eq.~(\ref{scalardefs}f) for $\chi_6^\pm$ and in the denominators 
of several terms in the inverse relations (\ref{inversescalar}). This factor vanishes at $r = 4.06$ for the 
trumpet geometry. More precisely, $(1 - 2\alpha e^{-4\varphi})$ is positive for $r < 4.06$ and negative for 
$r > 4.06$. Thus, as the pulse $\chi_6^-$ passes through $r = 4.06$, the factors $(1 - 2\alpha e^{-4\varphi})$
in its definition (\ref{scalardefs}f) change sign. This change of sign is responsible for the inversion 
that is seen in the incoming pulse $\chi_6^-$. Figure \ref{Dinversionfig} shows a time sequence of the inversion process. 
One can see from Figs.~\ref{Dmode9chi8at57} and \ref{Dmode9chi8at80} that a similar inversion occurs in the reflected pulse 
$\chi_6^+$ as it travels outward and passes through $r = 4.06$.

\subsection{Incident mode $\chi_4^-$}
\begin{figure}[htb] 
	{\includegraphics[scale = 1.4,viewport = 80 50 200 180]{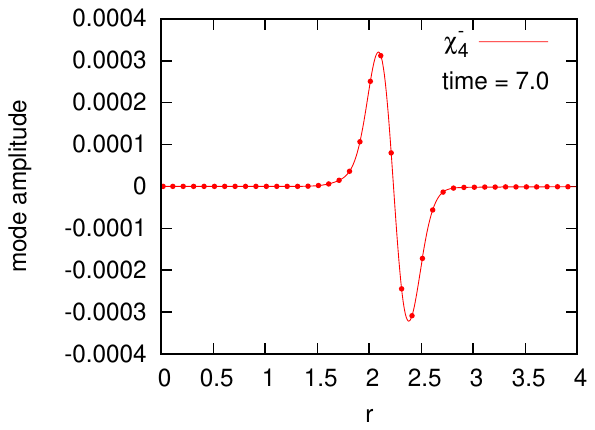}}
	\caption{Incoming mode $\chi_4^-$ at time $7.0$. (The initial excitation is in 
	 $\chi_4^-$.)}
	 \label{Dmode5chi5at35}
\end{figure}

\begin{figure}[htb] 
	{\includegraphics[scale = 1.4,viewport = 80 50 200 180]{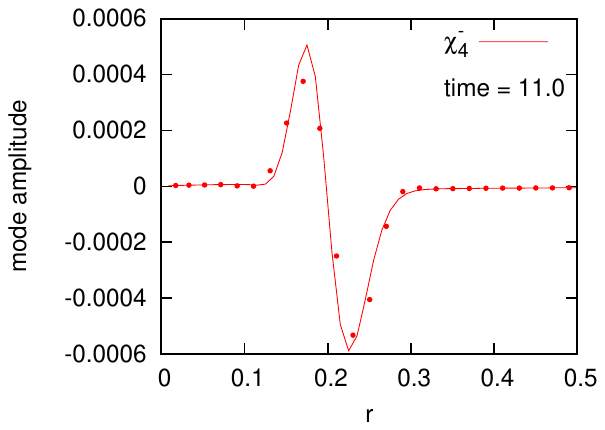}}
	\caption{Incoming mode $\chi_4^-$ at time $11.0$.  The dots coincide with 
	all of the data points for the low resolution simulation. (The initial excitation is in 
	 $\chi_4^-$.)}
	\label{Dmode5chi5at55}
\end{figure}

\begin{figure}[htb] 
	{\includegraphics[scale = 1.4,viewport = 80 50 200 180]{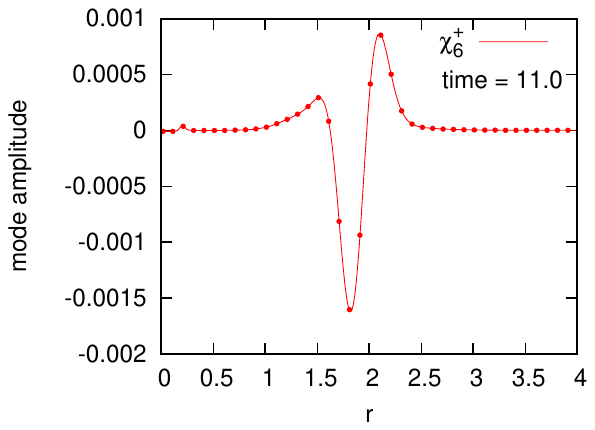}}
	\caption{Outgoing mode $\chi_6^+$ at time $11.0$. (The initial excitation is in 
	 $\chi_4^-$.)}
	 \label{Dmode5chi8at55}
\end{figure}

\begin{figure}[htb] 
	{\includegraphics[scale = 1.4,viewport = 80 50 200 180]{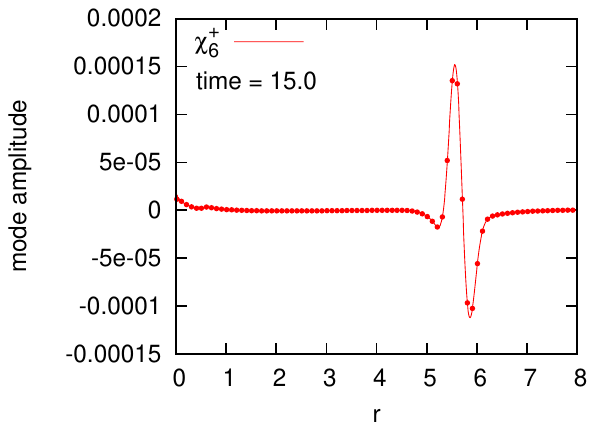}}
	\caption{Outgoing mode $\chi_6^+$ at time $15.0$. (The initial excitation is in 
	 $\chi_4^-$.)}
	 \label{Dmode5chi8at75}
\end{figure}

Figures \ref{Dmode5chi5at35} and \ref{Dmode5chi5at55} show the initial pulse $\chi_4^-$ propagating inward at times 
$t = 7.0$ and $t=11.0$, respectively. Figures \ref{Dmode5chi8at55} and \ref{Dmode5chi8at75} show a 
reflection in $\chi_6^+$ propagating outward at times $t=11.0$ and $t=15.0$, respectively. 
All of these figures show nice convergence, with no resolution--dependent time delay. 

These results require further analysis and explanation. Note that the pulse $\chi_6^+$, shown 
in Fig.~\ref{Dmode5chi8at55}, emerges from the puncture region {\em before} the 
incident pulse $\chi_4^-$ reaches the puncture boundary. Figure \ref{Dmode5chi5at55} 
shows that at time $t=11.0$, the incident pulse is at $r\approx 0.2$. Evidently the 
perturbation in the outgoing mode $\chi_6^+$ is caused by something other than  
a reflection of $\chi_4^-$ from the puncture boundary. 

\begin{figure}[htb] 
	{\includegraphics[scale = 1.4,viewport = 80 50 200 180]{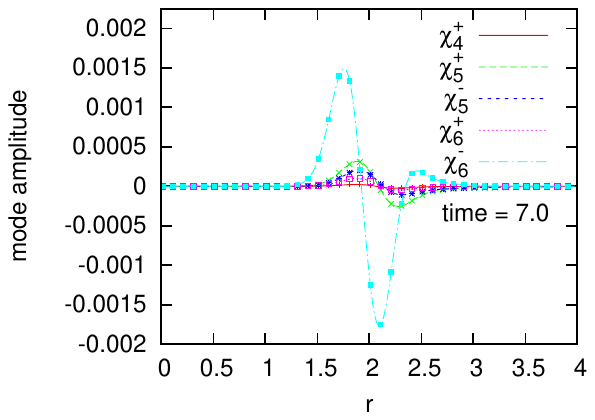}}
	\caption{Modes $\chi_4^+$, $\chi_5^\pm$, and $\chi_6^\pm$  at time $7.0$. 
	These modes are carried inward along with the initially excited mode $\chi_4^-$. 
	Compare with Fig.~\ref{Dmode5chi5at35}.}
	\label{Dmode5chiallat35}
\end{figure}

In Fig.~\ref{Dmode5chiallat35} we plot the  modes $\chi_4^+$, $\chi_5^\pm$, and 
$\chi_6^\pm$ at time $t = 7.0$. What is immediately clear from this figure is that 
the nonlinear interactions have produced a large  excitation in $\chi_6^-$. 
Observe that even with the ``gauge system" described in Sec.~III.D, the spatial variations 
in the  characteristic speeds will introduce a source in the evolution equation for $\chi_6^-$ that 
depends on $\chi_4^-$. This is a particularly strong coupling; note that 
the amplitude of $\chi_6^-$ 
at $t = 7.0$ is several times larger than the amplitude of $\chi_4^-$. 
(Recall, however, that relative amplitudes depend on the normalization 
of the modes.) 

Table I shows that near the puncture, mode $\chi_6^-$ propagates more quickly than mode 
$\chi_4^-$. As a consequence the excitation in $\chi_6^-$ races ahead of $\chi_4^-$. 
One can see this already at time $t=7.0$; $\chi_6^-$ is centered 
at $r\approx 1.9$ while $\chi_4^-$ is centered at $r \approx 2.2$. 
The large perturbation in $\chi_6^-$ reaches the puncture boundary at $t \approx 9.0$ 
and produces the large reflection in mode $\chi_6^+$, seen in Figs.~\ref{Dmode5chi8at55} 
and \ref{Dmode5chi8at75}. 

By time $t = 11.0$, the incident pulse $\chi_4^-$ has reached $r \approx 0.5$. 
Observe that at $t = 11.0$, Fig.~\ref{Dmode5chi8at55} shows a small hump in 
$\chi_6^+$ near $r \approx 0.5$. This hump is caused by the coupling between 
modes $\chi_4^-$ and $\chi_6^+$. The tail of this hump is seen in Fig.~\ref{Dmode5chi8at75} as an upturn in the 
data near the origin at $t=15.0$. 

\begin{figure}[htb] 
	{\includegraphics[scale = 1.4,viewport = 80 50 200 180]{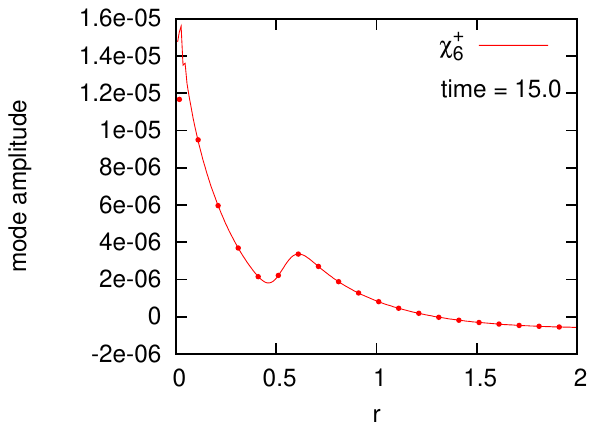}}
	\caption{Close--up view of outgoing mode $\chi_6^+$ at time $15.0$. 
	Compare with Fig.~\ref{Dmode5chi8at75}. (The initial excitation is in $\chi_4^-$.)}
	\label{Dmode5chi8at75close}
\end{figure}

Figure \ref{Dmode5chi8at75close} shows 
a close--up view of $\chi_6^+$ near $r=0$ at $t=15.0$.  
This view reveals another feature, a bump in $\chi_6^+$ 
that peaks around $r\approx 0.6$. Close inspection of the data shows that this bump has the following 
origin: The main reflection in $\chi_6^+$ creates an excitation in mode $\chi_5^-$, which then 
propagates inward toward the origin. In turn, this incoming wave  $\chi_5^-$ creates a 
distortion in $\chi_6^+$ that is carried inward along with $\chi_5^-$. It is this 
distortion that appears as the small bump in Fig.~\ref{Dmode5chi8at75close}. 

\begin{figure}[htb] 
	{\includegraphics[scale = 1.4,viewport = 80 50 200 180]{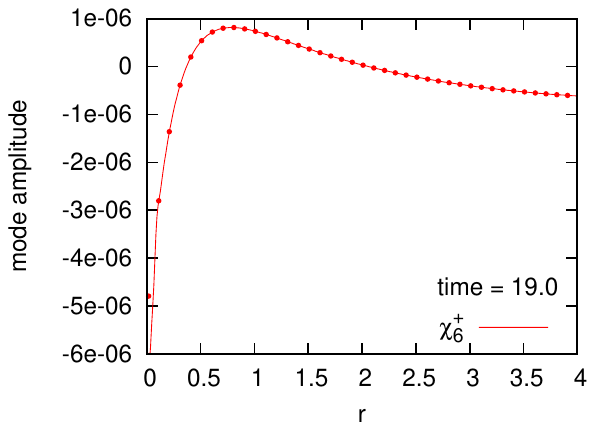}}
	\caption{Outgoing mode $\chi_6^+$ at time $19.0$. 
	 (The initial excitation is in $\chi_4^-$.)}
	\label{Dmode5chi8at95}
\end{figure}

The incident perturbation $\chi_4^-$  reaches the puncture boundary at about $t\approx 15.0$. 
It does not produce a significant reflection in mode $\chi_6^+$. 
Figure \ref{Dmode5chi8at95} shows the data for $\chi_6^+$ at $t=19.0$, when such a reflection would 
be expected to appear in the range $0 < r < 4.0$. By this time the excitation in $\chi_6^+$ has 
decayed to  
a ``tail" near the boundary that oscillates as it flattens to zero. Note that the 
amplitude in Fig.~\ref{Dmode5chi8at95} is 
quite small compared to the earlier reflections shown in Figs.~\ref{Dmode5chi8at55} 
and \ref{Dmode5chi8at75}. More 
importantly, the results are convergent: there are no resolution--dependent features. 

\subsection{Incident mode $\chi_5^-$}
\begin{figure}[htb] 
	{\includegraphics[scale = 1.4,viewport = 80 50 200 180]{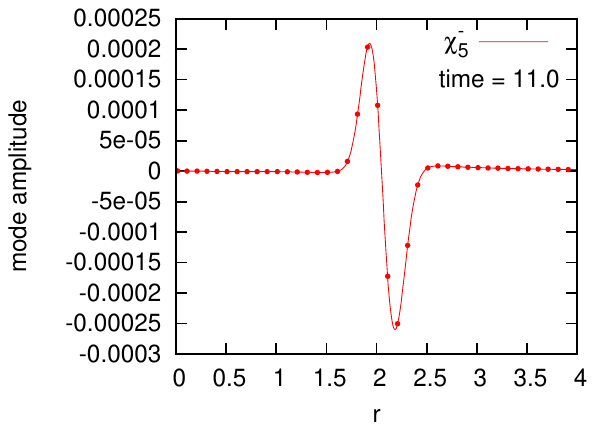}}
	\caption{Incoming mode $\chi_5^-$  at time $11.0$. (The initial excitation is in 
	 $\chi_5^-$.)}
	 \label{Dmode7chi7at55}
\end{figure}

Figure \ref{Dmode7chi7at55} shows the mode $\chi_5^-$ at time $t=11.0$ 
as it propagates inward toward the 
puncture boundary. Nonlinear interactions cause the other modes to 
become excited. By $t=7.0$, mode $\chi_6^-$ has been excited to 
an amplitude of about $A\approx 1\times 10^{-5}$. This mode moves more quickly 
than $\chi_5^-$ and reaches the puncture at $t\approx 9.0$. 
\begin{figure}[htb] 
	{\includegraphics[scale = 1.4,viewport = 80 50 200 180]{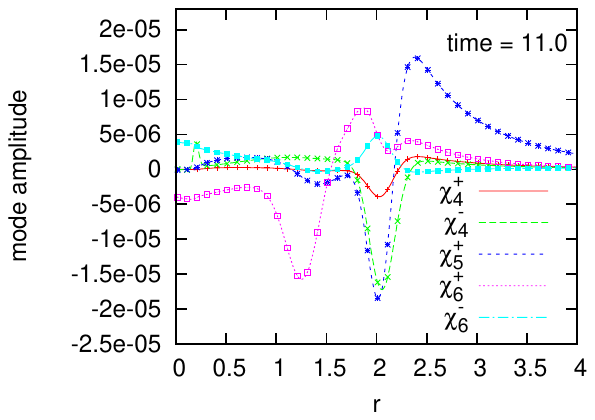}}
	\caption{Modes $\chi_4^\pm$, $\chi_5^+$, and $\chi_6^\pm$ at time $11.0$. 
	The mode $\chi_6^+$ is moving outward. The other modes are moving 
	inward along with the initial perturbation in $\chi_5^-$.}
	\label{Dmode7chiallat55}
\end{figure}

The modes $\chi_4^\pm$, $\chi_5^+$, and $\chi_6^\pm$ are 
plotted in Fig.~\ref{Dmode7chiallat55} at time $t=11.0$. 
The relatively large perturbation in $\chi_6^+$, with amplitude $A\approx 1\times 10^{-5}$, comes from 
the earlier reflection of $\chi_6^-$. This mode is moving outward at $t=11.0$, and  is 
convergent. 
The other modes shown in Fig.~\ref{Dmode7chiallat55} are moving inward, being carried by 
the initial perturbation $\chi_5^-$. 

\begin{figure}[htb] 
	{\includegraphics[scale = 1.4,viewport = 80 50 200 180]{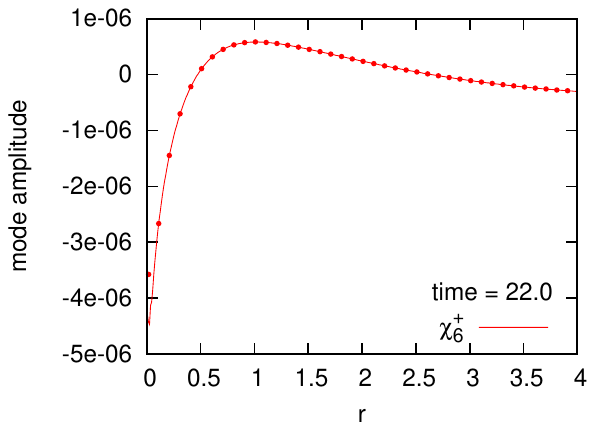}}
	\caption{Outgoing mode $\chi_6^+$  at time $22.0$. (The initial excitation is in 
	 $\chi_5^-$.)}
	 \label{Dmode7chi8at110}
\end{figure}

The main pulse $\chi_5^-$ reaches the puncture boundary at about $t\approx 20.0$. 
It does not produce a distinct reflection in mode $\chi_6^+$. 
Figure \ref{Dmode7chi8at110} shows mode $\chi_6^+$ at $t=22.0$, when such a reflection 
would appear in the range $0 < r < 4.0$. The only excitation 
in $\chi_6^+$ at this time is a decaying tail, similar to the profile seen at late 
times when the initial perturbation is in $\chi_4^-$; see Fig.~\ref{Dmode5chi8at95}. This tail 
is convergent, showing no resolution dependence. 

\subsection{Incident modes $\chi_1$, $\chi_2$ and $\chi_3$}
The results of these three cases are qualitatively very similar to one another. The incident 
pulse takes about $t\approx 45$ to propagate from its initial location at $r_0 = 4.0$ to the 
puncture boundary. Long before that, the nonlinear interactions create excitations in 
the other modes. In particular, the excitation in mode $\chi_6^-$ races ahead and reaches the 
puncture boundary at $t\approx 4$. As expected, this produces a reflection in mode 
$\chi_6^+$ that propagates outward to the black hole exterior. When the initial perturbations 
in $\chi_1$, $\chi_2$ or $\chi_3$ hit the puncture boundary, they do not appear to 
couple to any other modes. At late times   we only see a residual tail 
in $\chi_6^+$ that decays in time. 
\begin{figure}[htb] 
	{\includegraphics[scale = 1.4,viewport = 80 50 200 180]{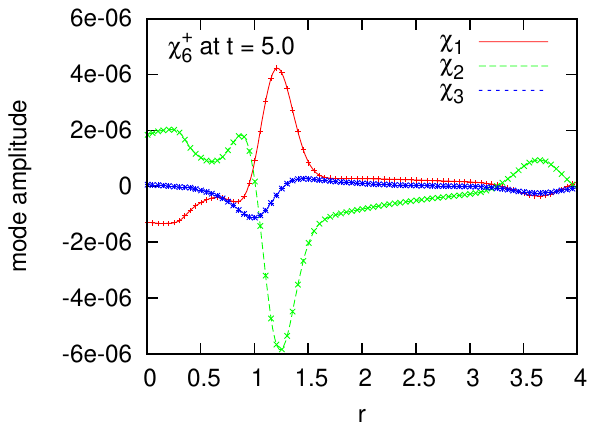}}
	\caption{Mode $\chi_6^+$ at time $5.0$. The three curves correspond 
	to initial excitations in $\chi_1$, $\chi_2$ and $\chi_3$.}
	 \label{mode8attime5}
\end{figure}
\begin{figure}[htb] 
	{\includegraphics[scale = 1.4,viewport = 80 50 200 180]{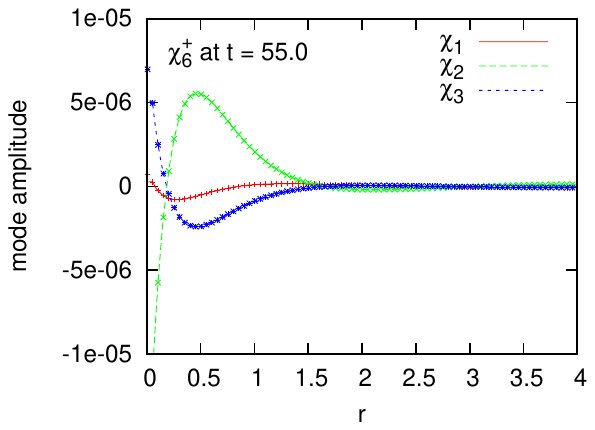}}
	\caption{Mode $\chi_6^+$  at time $55.0$. The three curves correspond 
	to initial excitations in $\chi_1$, $\chi_2$ and $\chi_3$.}
	 \label{mode8attime55}
\end{figure}

Figure \ref{mode8attime5} shows $\chi_6^+$ for the three cases with initial excitations 
in $\chi_1$, $\chi_2$ and 
$\chi_3$ at time $t = 5.0$. The large pulses near $r = 1.2$ are reflections from $\chi_6^-$. These
wave forms are traveling 
outward with positive speed. The smaller pulses near $r=3.6$ are excitations that are 
carried inward along with the initial perturbations in $\chi_1$, $\chi_2$ or $\chi_3$. 
None of these curves show any resolution dependence. As the initial perturbation passes through the 
puncture boundary, the  mode $\chi_6^+$ develops a relatively large amplitude profile. That profile 
decays in time, and does not appear to propagate outward. 
In Fig.~\ref{mode8attime55}
we see $\chi_6^+$ at  time $t=55$, after the initial pulse has hit the puncture boundary. 
This figure shows the residual profile and no resolution dependence. 

\section{Interpretation and Discussion} 
Only the modes $\chi_6^+$ 
and $\chi_8^+$ are able to propagate from the puncture boundary to the black hole exterior. 
When starting the evolution from a wormhole configuration, modes $\chi_4^+$, $\chi_5^+$,
$\chi_9^+$ and $\chi_{10}^+$ 
are outgoing at the puncture boundary. However, these modes quickly acquire negative speeds near the 
boundary and their characteristic curves never extend from the puncture boundary to the black hole 
exterior. 

The modes $\chi_6^\pm$ and $\chi_8^\pm$ have constant speeds, $\pm \sqrt{g^{rr}} \approx \pm 1$ and 
$\pm \sqrt{3g^{rr}}/2 \approx \pm 0.87$ respectively, 
near the puncture boundary of a trumpet geometry. This implies that these modes propagate between the 
puncture boundary and the black hole exterior with no resolution dependence. 
These modes do not recognize that the puncture boundary moves when the resolution is changed. 
In effect, {\em modes $\chi_6^\pm$ and $\chi_8^\pm$ propagate in the conformal geometry, 
not the physical geometry.} 

The other modes, namely $\chi_1$, $\chi_2$, $\chi_3$, $\chi_4^\pm$, $\chi_5^\pm$, $\chi_7$, 
$\chi_9^\pm$ and $\chi_{10}^\pm$, 
have speeds $-\beta^r \approx -0.53r$ near the puncture. They propagate in the physical geometry. 
These modes are affected by the movement of the puncture boundary when the resolution is changed. 
Once the spatial slice settles to a trumpet geometry, all of these modes are incoming at 
the puncture boundary. 

The distinction between modes that propagate in the physical geometry and those that propagate in 
the conformal geometry can be understood by examining the speeds from Tables I and II. 
The coordinate
speeds for modes $\chi_6^\pm$ and $\chi_8^\pm$ are built from the shift vector and the 
conformal metric component $g^{rr}$. The physical metric does not appear. On the other hand, 
the coordinate speeds for modes $\chi_4^\pm$, $\chi_5^\pm$, $\chi_9^\pm$ and $\chi_{10}^\pm$ 
are built from the shift vector and the physical metric component $e^{-4\varphi}g^{rr}$. 

The time required for a perturbation in any mode to travel between the puncture 
boundary at $r \approx h/2$ and a finite radius $r_0$ is 
\begin{equation}
	T \approx \int_{h/2}^{r_0} \frac{dr}{|v|} \ ,
\end{equation} 
where $v$ is the coordinate speed. For the modes that propagate in the conformal geometry, 
$v$ is constant near the puncture boundary and $T$ depends linearly on the grid resolution $h$. 
In the limit $h\to 0$, this dependence drops out. In this sense, the propagation time does 
not depend on resolution. For the modes that propagate in the physical geometry, $v$ 
behaves like $|v|\sim r$ near the puncture boundary of a trumpet. For these modes 
$T$ diverges like $|\ln(h)|$. As the resolution is increased  $h$ is decreased and the 
puncture boundary is pushed farther down the trumpet throat. 
The propagation time $T$ goes to infinity. 

Once the wormhole  has evolved into a trumpet, all of the modes that propagate in the physical geometry 
have negative speeds at the puncture boundary. The concern is that a disturbance in one of these modes 
could propagate inward to the puncture boundary where it might couple to one of the 
outgoing superluminal modes $\chi_6^+$ or $\chi_8^+$. Since the time required for the initial 
disturbance to reach the puncture boundary is resolution dependent, the reflection would 
be delayed in time as the resolution is increased. 

The numerical experiments in Sec.~V suggest that there is no coupling at the puncture boundary between the
modes that propagate in the physical geometry and the modes that propagate in the conformal geometry. 
These experiments were limited 
to spherical symmetry, so not all possible couplings could be tested. 

The lack of coupling between the two types of modes can be understood by considering the 
finite differencing stencil  as it appears from the perspectives of the 
conformal and physical geometries. From the point of view of the physical geometry, the 
finite differencing stencil is one sided at the puncture boundary. We can see this in 
Fig.~\ref{CoordsOnCylinder}.  For any point on the puncture boundary, the legs 
of a stencil will always extend outward (into the computational domain). 
For the modes that travel in the physical geometry, the puncture method correctly avoids placing any 
boundary conditions at the puncture boundary by using one--sided stencils. These modes are 
simply advected through the puncture boundary and off the computational domain. 

For the modes that travel in the conformal geometry, the finite differencing stencils appear 
as standard stencils that surround the origin of a computational grid in  $R^3$. 
For these modes, there is no boundary and the puncture method correctly treats the puncture 
like any other point in the computational domain. 

Because the modes $\chi_6^\pm$ and $\chi_8^\pm$ propagate in the conformal geometry, which is 
smooth at the origin, we expect the puncture method to impose some type of smoothness conditions 
at $r=0$. It is not entirely clear what those conditions might be since  
the unperturbed fields are not all smooth at that point. The following observations are suggestive. 
If we assume that the perturbations in the vector fields $B^a$ and $\beta^a$  
are smooth at the origin, then we must have $\tilde B^n = 0$ and $\tilde\beta^n = 0$.
From Eq.~(\ref{scalardefs}f) we see that the modes $\chi_6^\pm$ obey 
$\chi_6^\pm(0) = \mp \partial_n \tilde\beta^n(0) $
at $r=0$ for a trumpet geometry. Thus, we expect that the relation 
\begin{equation}
	\chi_6^+(0) + \chi_6^-(0) = 0
\end{equation} 
should hold for all of the simulations in Sec.~V. A close inspection of the data 
shows that this is indeed the case. 

For BSSN with standard gauge conditions, some modes effectively propagate in the physical 
geometry and some effectively propagate in the conformal geometry. In light of this understanding, 
let us examine the scalar field example of Sec.~II. In that case the characteristic speeds 
are $v = \pm r/\sqrt{1+r^2} \approx \pm r$ near the puncture boundary. Both the incoming and 
outgoing modes propagate in the physical geometry. The time $T$ for these modes to travel between the puncture 
boundary and any finite location diverges like $|\ln(h)|$. Moreover, the finite difference 
stencil for both of these modes is one sided at the puncture boundary. This is bad. For a properly 
constructed numerical code, the outgoing mode (the mode that is propagating into the 
computational domain) should be fixed by boundary conditions. This does not happen when we naively apply the 
puncture method as in Sec.~II. Correspondingly, the code used in Sec.~II does not 
work---the simulations do not converge 
to a continuum solution as the resolution is increased. 

On the other hand, a properly constructed numerical code for BSSN with standard gauge should have 
precisely the properties of the puncture method. That is, the code should use something like 
one--sided finite difference stencils for the modes that propagate in the physical geometry, since 
all of these modes are incoming at the puncture boundary. For the modes that propagate in the 
conformal geometry there is no boundary, so the code should use the same stencils near the 
origin as elsewhere. 

The analysis in this paper assumes that the gamma--driver shift 
condition includes all advection terms. There are three such terms, one in Eq.~(\ref{BSSNeqns}g) and 
two in Eqs.~(\ref{BSSNeqns}h). Let me comment on the cases in which 
one or more advection terms are omitted. What we look for is a case with an 
outgoing mode whose coordinate speed goes to zero at the puncture. To determine the speeds near the puncture 
boundary, I will assume that the geometry evolves to a stationary trumpet 
with $\alpha \approx ar$, $\beta^r \approx br$, $g^{rr} \approx g$ and $\varphi \approx -\ln(pr)/2$ near 
the puncture boundary, where 
$a$, $b$, $g$ and $p$ are all constants. This assumption holds if we include all advection terms, 
and it holds if we include no advection terms. 

With the assumption above, there is only one scalar mode and one vector mode 
whose speeds can approach zero through positive 
values near $r=0$. This happens when the second advection term in Eq.~(\ref{BSSNeqns}h) is 
dropped but the other advection terms are kept. [The second advection term is contained in  
$(\partial_t\Lambda^a)_{\rm rhs}$.] In that case the scalar mode has speed $b^3 r^3/g$ 
and the vector mode has speed $4b^3 r^3/(3g)$ near the puncture boundary. If either of 
these modes is excited at the puncture boundary, it will propagate to the black hole 
exterior with a  resolution--dependent time delay. More precisely, the propagation 
time will diverge like $1/h^2$ as the resolution is increased. The conclusion is that 
we should not drop the second advection term in the gamma--driver shift equation 
Eq.~(\ref{BSSNeqns}h) unless 
one or both of the other advection terms are dropped as well. 

%
\appendix
\section{Cartoon BSSN code}
\begin{figure}[htb] 
	{\includegraphics[scale = 0.8]{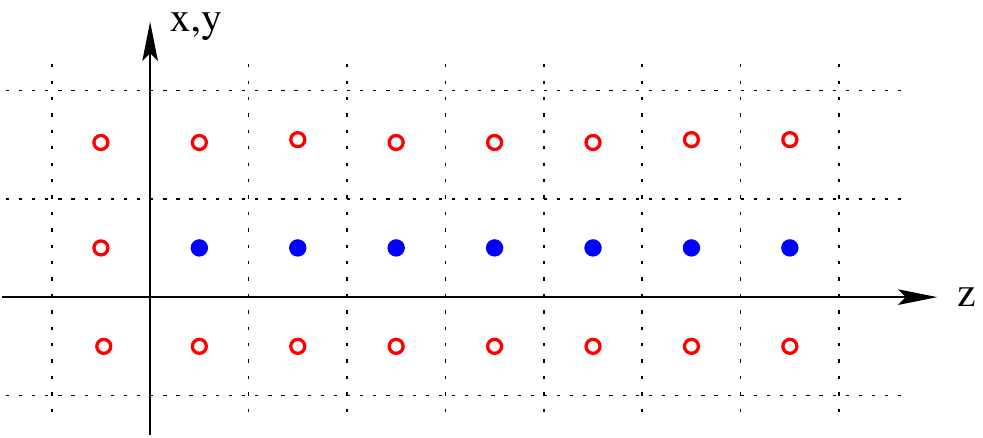}}
	\caption{Grid for the cartoon code. The physical grid points 
	are filled dots, and  buffer points are open circles.}
	\label{cartoongrid}
\end{figure}

The cartoon code evolves the BSSN plus standard gauge system in spherical symmetry
using Cartesian finite difference stencils. The computational grid is shown in Fig.~\ref{cartoongrid}. 
The puncture is at the origin. The physical grid consists of the single line of 
grid points at $x/h = y/h = 1/2$ and $z/y = 1/2, 3/2, \ldots$, where $h$ is the grid 
spacing. These points are shown as solid dots in the figure. 
The open circles represent buffer points, which are needed for computing 
finite difference derivatives. The figure shows one layer of buffer points; 
the actual code uses two layers of buffer points. 

The buffer points are filled by spherical symmetry using known values of the fields at  
physical grid points. Consider a scalar field such 
as the lapse function $\alpha$. The value of $\alpha$ at a buffer point $(x_g,y_g,z_g)$
is found by considering the coordinate distance from the origin: $r_g = \sqrt{x_g^2 + y_g^2 + z_g^2}$. 
We compute the  
value of $z$ along the physical grid line for the point at radius $r_g$. That is, 
we set $\sqrt{(h/2)^2 + (h/2)^2 + z^2} = r_g$ and solve for $z$; this gives  
\begin{equation}\label{zvalue}
	z = \sqrt{x_g^2 + y_g^2 + z_g^2 - h/2} \ .
\end{equation}
I use an eighth--order interpolation along the $z$ direction to find $\alpha$ at radius $r_g$. 

As an example, consider the buffer point $(5h/2,3h/2,9h/2)$. According to Eq.~(\ref{zvalue}) 
we have $z = 5.315h$. Thus, the point $(h/2,h/2,5.315h)$ lies along the physical 
grid line at the same radius as the buffer point.  Note that $z = 5.315h$ lies between 
$9h/2$ and $11h/2$. The
value of $\alpha$ at  $(h/2,h/2,5.315h)$ is found by interpolation over the 
eight physical points with $z/h = 3/2,\ldots 17/2$. This is the value assigned to $\alpha$ 
at the buffer point. 

The conformal connection $\Lambda^a$, the shift $\beta^a$, and the auxiliary field $B^a$ are 
all spatial vectors. Let $V^a$ denote one of these vectors. In spherical symmetry, 
the Cartesian components of any vector field can be written as 
\begin{equation}\label{vectorss}
	V^a = S(r) \frac{x^a}{r} \ ,
\end{equation}
where $S(r)$ is a scalar field. Here, $x^a$ are Cartesian coordinates and $r = \sqrt{x^2 + y^2 + z^2}$. 
Relation (\ref{vectorss}) can be inverted, 
\begin{equation}\label{invvectorss}
	S = V^a \frac{x_a}{r}  \ ,
\end{equation}
where indices on $x^a$ are lowered with the flat Cartesian metric $\delta_{ab}$. 
In filling buffer points for vectors, we first apply Eq.~(\ref{invvectorss}) to compute the 
scalar $S$. Next, we apply the scheme outlined above to fill the buffer points for $S$. 
Finally, the buffer points for $V^a$ are computed from Eq.~(\ref{vectorss}). 

Let $T_{ab}$ denote one of the symmetric tensor fields $g_{ab}$ or $A_{ab}$. 
In spherical symmetry we have 
\begin{equation} \label{tensorss}
	T_{ab} = P(r) \frac{x_a x_b}{r^2} + Q(r) \delta_{ab} \ ,
\end{equation}
where $P$ and $Q$ are scalars. 
This relation can also be expressed as 
\begin{equation}
	T_{ab}dx^a dx^b = (P+Q)dr^2 + r^2 Q d\Omega^2 \ ,
\end{equation}
where $d\Omega^2$ is the metric on the unit sphere. 
Buffer points for $T_{ab}$ are filled by first computing the scalars from the inverse relations 
\begin{subequations}\label{invtensorss}
\begin{eqnarray}
	P + 3Q & = & T_{ab}\delta^{ab} \ ,\\
	 P + Q & = & T_{ab} \frac{x^a x^b}{r^2}  \ .
\end{eqnarray}
\end{subequations}
We then fill the buffer points for the scalars, 
as outlined above. Finally we apply Eq.~(\ref{tensorss}) to obtain buffer point 
values for $T_{ab}$. 

Note that the calculations discussed here rely on the fact that 
all of the fields transform as simple tensors with no density weights. 
This is the case for the covariant formulation of BSSN and the standard gauge
\cite{Brown:2009dd}. 

The cartoon BSSN code uses fourth--order Runge--Kutta for time integration. It uses 
standard fourth--order centered finite difference stencils for spatial 
derivatives, like those displayed in Eqs.~(\ref{fourthorderstencils}). 
The only exceptions to this rule are the advection 
terms, which have the form $\beta^a\partial_a F$ for a tensor or tensor component $F$. These 
terms are upwinded. The code as it is now written includes two layers of buffer points. This limits the 
size of the finite difference stencils such that the advection terms are only third--order accurate. 
For example, the finite difference approximation to $\beta^z \partial_z F$ with $\beta^z>0$ is
\begin{equation}
	(\beta^z\partial_z F)_k = \frac{1}{6h}\beta^z_k(-2F_{k-1} - 3F_k + 6F_{k+1} - F_{k+2})
\end{equation}
where $k$ labels the grid points and $h$ is the grid spacing. 
The overall accuracy of the cartoon code is limited to third 
order by the presence of these advection terms. 

\begin{figure}[htb] 
	{\includegraphics[scale = 1.4,viewport = 80 50 200 180]{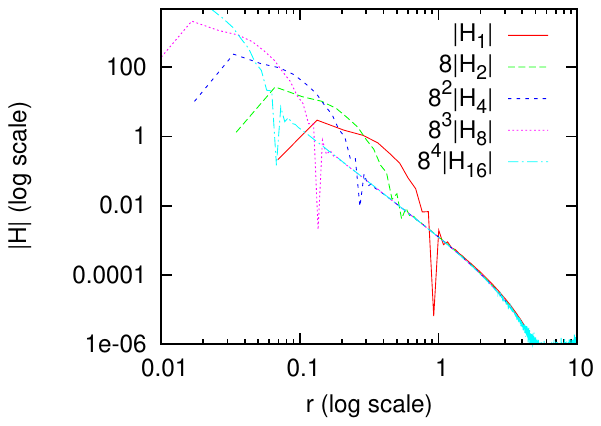}}
	\caption{Absolute value of the Hamiltonian constraint at five different resolutions. 
	The curves are scaled by powers of $8$ to show third--order convergence.}
	\label{Hconvergence}
\end{figure}

\begin{figure}[htb] 
	{\includegraphics[scale = 1.4,viewport = 80 50 200 180]{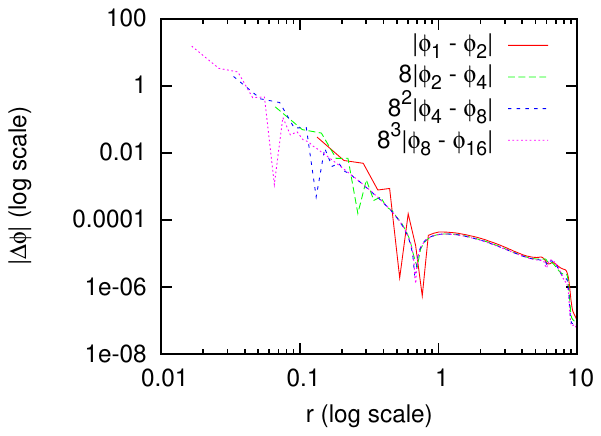}}
	\caption{Absolute value of the difference between the conformal 
	factor at successive resolutions. The curves are scaled by powers of $8$ to show 
	third--order convergence.}
	\label{phiconvergence}
\end{figure}

Figures \ref{Hconvergence} and \ref{phiconvergence} show the results of convergence 
tests for the cartoon code. The data for both of these figures come from a 
simulation in which a wormhole slice of Schwarzschild  
evolves into a trumpet geometry. The data are taken at time $t=10$. At this time the various 
fields, including the lapse function and 
shift vector, have settled fairly close to their trumpet values in the immediate vicinity of the 
puncture (say, $r \lessthansimilarto 0.1$) but are still changing rapidly away from the puncture 
($r\greaterthansimilarto 0.1$). 

Simulations at five different grid resolutions were used for these tests. 
Let us label these resolutions by integer subscripts 
$1$, $2$, $4$, $8$, and $16$. The grid spacings are $h_1 = 1/12.5$, $h_2 = 1/25$, 
$h_4 = 1/50$, $h_8 = 1/100$, and $h_{16} = 1/200$. 

Figure \ref{Hconvergence} shows the absolute value of the Hamiltonian constraint. 
The curves in this figure are scaled by factors of $8$ at successive resolutions, 
appropriate for a third--order numerical scheme. As one can see, the curves 
overlap in the limit of increasing resolution. This shows that the code is indeed 
third--order accurate. 

Figure \ref{phiconvergence} shows the result of a three--point convergence test 
for the conformal factor $\varphi$. Each curve is obtained from the difference between the conformal factor at 
successive resolutions, scaled by an appropriate power of $8$. 
The curves in this plot also overlap in the limit of increasing resolution. This confirms 
that the code is third--order convergent. 

These results require some elaboration. It has been stated in a number of publications
that puncture evolution codes do not converge near the puncture.
What we see from these figures is that the errors  are large near the puncture. This does 
not imply a lack of convergence. A code is convergent if, at any given point 
in the computational domain, the scaled errors coincide with one another in the limit of high resolution. For 
example, consider the point $r = 0.3$ on the graph of Fig.~\ref{Hconvergence}. This point is 
about half way between $0.1$ and $1$ on the logarithmic scale. 
The two lowest resolution curves do not coincide with the higher resolution curves at $r = 0.3$, due to 
the presence of large finite differencing errors. However, the three higher resolution curves 
overlap nicely. Thus, the code appears to be third--order convergent at $r = 0.3$. 

Similarly, let us consider the point $r = 0.1$. None of the 
curves overlap at this point. This is due to insufficient resolution. It is fairly clear from 
examining the sequence of curves that if we were to add a higher resolution run, 
with $h_{32} = 1/400$, that the curves $8^4|{\cal H}_{16}|$ and $8^5|{\cal H}_{32}|$ would closely agree. 
We then expect that at $r=0.1$ the code is third--order convergent. In fact, the code  appears to be 
third--order convergent at all points $r>0$. This is as it should be since the data are smooth everywhere except 
at $r=0$. A properly constructed code should be convergent at all points in the 
computational domain, excluding only the puncture points. 

\section{Trumpet geometry in Gamma--driver coordinates}
In Sec.~V we considered the evolution of perturbations on a trumpet slice 
of a single Schwarzschild black hole. The perturbations were defined 
in terms of characteristic fields through Eqs.~(\ref{inversescalar}). This appendix is 
devoted to a description of the unperturbed trumpet geometry. 

In this paper, the term ``trumpet geometry" refers to a stationary 1+log slice of 
the Schwarzschild geometry. Such a slice can be expressed in various spatial coordinate 
systems. One technique for finding a trumpet slice in isotropic coordinates 
has been described elsewhere \cite{Hannam:2008sg,Bruegmann:2009gc}.
What is needed for the analysis in Sec.~V is a trumpet slice in stationary gamma--driver coordinates. 
That is, the data should satisfy the gamma--driver shift equations (\ref{BSSNeqns}g,h) including 
advection terms and with time derivatives set to zero:
\begin{subequations}\label{stationaryGDS}
\begin{eqnarray}
	0 & = & \beta^c \mathring D_c \beta^a + \frac{3}{4} B^a \ ,\\
	0 & = & \beta^c \mathring D_c B^a - \beta^c \mathring D_c \Lambda^a  \ .
\end{eqnarray}
\end{subequations}
The damping parameter $\eta$ was set to zero in all simulations presented in this paper; 
it is set to zero here as well. 

Let me begin by reviewing the construction of trumpet data in isotropic coordinates, then show 
how the results can be extended to stationary gamma--driver coordinates. 
A stationary, spherically symmetric geometry can be described 
by the metric 
\begin{equation}\label{metricell}
	ds^2 = -(\alpha^2 + \beta^2)dt^2 + 2\beta\,d\ell\,dt + d\ell^2 
	+ R^2 d\Omega^2
\end{equation} 
where the lapse $\alpha$, shift $\beta$, and areal radius $R$ are functions 
of the proper distance coordinate $\ell$. Recall that 
$d\Omega^2 \equiv  d\theta^2 + \sin^2\theta\,d\phi^2$ is the line element for the 
unit sphere. The Einstein equations imply 
\begin{equation}\label{einsteineqns}
	\alpha = R' \ ,\qquad \alpha^2 - \beta^2 = 1 - 2M/R \ ,
\end{equation}
where $M$ is an integration constant and the prime denotes $\partial/\partial\ell$. 
The 1+log slicing condition (\ref{BSSNeqns}f), along with stationarity, gives
\begin{equation}\label{alphaprime}
 	 \alpha' = \frac{2\beta'}{\beta} + \frac{4R'}{R} \ .
\end{equation} 
Equations~(\ref{einsteineqns}) can be used to eliminate $\alpha$ and $\beta$ 
from Eq.~(\ref{alphaprime}); 
this yields 
\begin{equation}\label{Rdoubleprime}
	R'' = \frac{2R'[3M - 2R + 2R{R'}^2]}{R[2M-R-2RR' + R{R'}^2]} \ .
\end{equation}
In Ref.~\cite{Hannam:2006vv}, Hannam {\em et al.} argue that the numerator and denominator on 
the right--hand side of Eq.~(\ref{Rdoubleprime}) must vanish separately at some particular value of $\ell$. 
Solving these conditions simultaneously shows that the equations 
$R' = -3 + \sqrt{10}$ and $M/R = 4(-3 + \sqrt{10})$ must hold for some value of $\ell$. 

Equation (\ref{alphaprime}) can be integrated to $\beta R^2 = \sqrt{C}e^{\alpha/2}$, 
where $C$ is an integration constant. By using Eqs.~(\ref{einsteineqns}) we can 
rewrite this expression in terms of $R$ and $R'$: 
\begin{equation}\label{Requation}
	C e^{R'} = R^4 [{R'}^2 - 1 + 2M/R] \ .
\end{equation}
The values previously obtained for $R$ and $R'$ at one value of $\ell$ can be used to solve 
for the constant, with the result  
\begin{equation}
	C = \frac{M^4 e^{3-\sqrt{10}}}{128(\sqrt{10}-3)^3} \approx 1.55\,M^4  \ .
\end{equation}
Note that in the limit as $\ell \to -\infty$, the trumpet geometry satisfies $R' \to 0$. Equation (\ref{Requation}) 
has two real solutions for $R$ in this limit,  but only one 
has $R'$ approaching $0$ through positive values. For that solution we find that the trumpet throat has 
radius $R \approx 1.312\,M$ in the limit $\ell \to -\infty$. 

The proper distance can be written as 
\begin{equation}\label{ellintegral}
	\ell = \int \frac{dR}{R'} \ ,
\end{equation}
where $R'$ is considered to be a function of $R$. This dependence, $R'$ as a function of $R$, is 
found by applying Newton's method to Eq.~(\ref{Requation}).
Then the integral (\ref{ellintegral}) can be evaluated numerically, starting from large negative $\ell$ 
where $R \approx 1.312\,M$. This gives $\ell$ as a function of $R$. 

Equivalently, the analysis above defines $R$ as a function of $\ell$. 
The data $\alpha$ and $\beta$, as functions of $\ell$, are found from Eqs.~(\ref{einsteineqns}). 
The components of the extrinsic curvature are computed from the spacetime 
metric (\ref{metricell}) as
\begin{subequations}
\begin{eqnarray}
	K_{\ell\ell} & = & \beta \left( \frac{R''}{2R'} - \frac{2}{R}\right) \ ,\\
	K_{\theta\theta} & = & \beta R \ ,\qquad K_{\phi\phi} = \beta R\sin^2\theta \ .
\end{eqnarray}
\end{subequations}
These are determined as functions of proper distance $\ell$ from the numerical solution $R(\ell)$. 

We now switch to isotropic coordinates, with spatial metric 
\begin{equation}\label{metricrho}
	ds^2 = \Psi^4(d\rho^2 + \rho^2 d\Omega^2 ) \ .
\end{equation}
Comparing with the spatial part of the metric (\ref{metricell}), we have $\Psi = \sqrt{R/\rho}$ and
$\partial \rho/\partial\ell = \rho/R$. Since $R$ is known as a function of $\ell$, this
second relation can be integrated numerically to give $\rho$ as a function of $\ell$. 
Turning this around, we can 
consider $\ell$ as a function of $\rho$. Then the conformal factor 
$\Psi$ is determined as a function of the isotropic coordinate $\rho$: explicitly,  
$\Psi = \sqrt{R(\ell(\rho))/\rho}$.  Likewise, the lapse function becomes a known function of the 
isotropic radius: $\alpha = \alpha(\ell(\rho))$. The
shift vector in isotropic coordinates is 
$\beta^\rho = (\rho/R)\beta$, and the radial--radial component of the extrinsic curvature is 
$K_{\rho\rho} = (R/\rho)^2 K_{\ell\ell}$. These are now determined as functions of $\rho$ 
since the dependence $\ell(\rho)$ is known. 

The stationary gamma--driver shift equations with $\eta = 0$ are 
written in Eqs.~(\ref{stationaryGDS}) above. 
Assuming $\beta^a \ne 0$, the second of these equations can be integrated to obtain 
$B^a = \Lambda^a + {\rm const}$. We can choose initial data such that 
$B^a = \Lambda^a$. Then the first of Eqs.~(\ref{stationaryGDS}) and 
the definition (\ref{Lambdadef}) for $\Lambda^a$ yield 
\begin{equation}\label{gammadrivercoords}
	0 = g^{bc}\left(\Gamma^a_{bc} - \mathring\Gamma^a_{bc}\right)  
		+ \frac{4}{3}\beta^c\mathring D_c\beta^a \ .
\end{equation}
Our goal is to find a change of spatial coordinates, from isotropic coordinates $\rho$, $\theta$, $\phi$ to 
``gamma--driver coordinates" $r$, $\theta$, $\phi$, such that Eq.~(\ref{gammadrivercoords}) is satisfied 
in the new coordinate system. 

We begin by rewriting the spatial metric (\ref{metricrho}) as 
\begin{equation}\label{metricr}
	ds^2 = \Psi^4 \left( \frac{\rho^2 \rho'}{r^2}\right)^{2/3} \left[ 
	\left( \frac{r \rho'}{\rho}\right)^{4/3} dr^2 + \left( \frac{r^2 \rho}{\rho'}\right)^{2/3} 
	d\Omega^2 \right] \ .
\end{equation}
Here, $\rho'$ is the derivative of $\rho$ with respect to the new radial coordinate $r$. 
Observe that the conformal part of this metric (the factor in square brackets) has determinant 
$r^4 \sin^2\theta$. With this 
choice the conformal metric will have determinant $1$ in Cartesian 
coordinates. Now use the conformal metric from Eq.~(\ref{metricr}) to write the gamma--driver shift 
equation (\ref{gammadrivercoords}) explicitly: 
\begin{eqnarray}\label{GDcoordexplicit}
	0 & = & \frac{1}{3r}\left( \frac{r\rho'}{\rho}\right)^{2/3} \left[ 6 -\frac{4\rho}{r\rho'} 
	- \frac{2\rho^2}{r^2{\rho'}^2} + \frac{4\rho^2 \rho''}{r{\rho'}^3} \right] 
	\nonumber \\  & & + \frac{4}{3} 
	\beta^\rho \left[  \frac{ {\beta^\rho}'}{\rho'} - \frac{ \beta^\rho \rho''}{{\rho'}^3}
	\right]  \ .
\end{eqnarray}
Here, the shift has functional dependence $\beta^\rho = \beta^\rho(\rho(r))$. 

Equation (\ref{GDcoordexplicit}) is a second order differential equation for $\rho$ as a function 
of $r$. We can turn this around and 
view it as an equation for $r$ as a function of $\rho$. In doing so, we must make the changes 
$\rho' = 1/r'$ and $\rho'' = - r''/{\rho'}^3$, where primes on $r$ denote derivatives with 
respect to $\rho$. 
With this view of Eq.~(\ref{GDcoordexplicit}), the shift has functional dependence 
$\beta^\rho = \beta^\rho(\rho)$. Recall that $\beta^\rho$ is known as a function of $\rho$ 
from the numerical trumpet solution in isotropic coordinates.  

The analysis above yields the following second order elliptic differential equation 
for $r$ as a function of $\rho$:
\begin{widetext}
\begin{equation}\label{ellipticeqnforr}
	r'' = \biggl[ (\rho r')^{7/3} + 2(\rho r')^{4/3} r - 3(\rho r')^{1/3} r^2  
		- 2\rho\beta^\rho {\beta^\rho}' {r'}^2 r^{7/3} \biggr] 
		\biggl[ 2 \rho{\beta^\rho}{}^2 r' r^{7/3} - 2(\rho^2 r')^{1/3} r \biggr]^{-1} \ .
\end{equation}
\end{widetext}
It is helpful to make the change of variables $\sigma \equiv r/\rho$. 
The resulting equation can be solved for  $\sigma$ as a function of $\rho$ with boundary 
conditions $\sigma = 1$ as $\rho \to 0$ and $\rho \to \infty$. 
Any number of techniques can be used;  I use a simple multigrid relaxation scheme. 
Once we have found $r$ as a function of $\rho$, the usual tensor transformation equations 
can be used to determine the lapse function, shift vector, and extrinsic curvature in the 
new coordinates $r$, $\theta$, and $\phi$. 

\begin{figure}[htb] 
	{\includegraphics[scale = 1.4,viewport = 80 50 200 180]{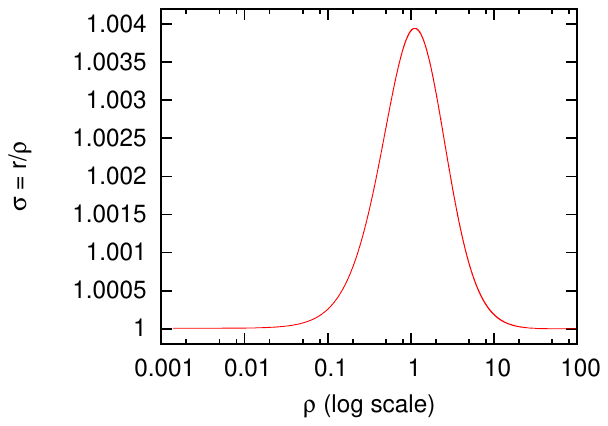}}
	\caption{The ratio $\sigma = r/\rho$, where $r$ is the radial coordinate for the 
	gamma--driver shift coordinates and $\rho$ is the isotropic radius.}
	\label{sigmavsrho}
\end{figure}

I have written a stand--alone numerical code to carry out the above calculations.
Figure \ref{sigmavsrho} is a plot of $\sigma$ versus $\rho$. It is immediately clear 
that $\sigma$ remains close to unity. Consequently 
the stationary gamma--driver coordinate $r$ differs by a relatively small amount from the 
isotropic radius $\rho$.

\begin{figure}[htb] 
	{\includegraphics[scale = 1.4,viewport = 80 50 200 180]{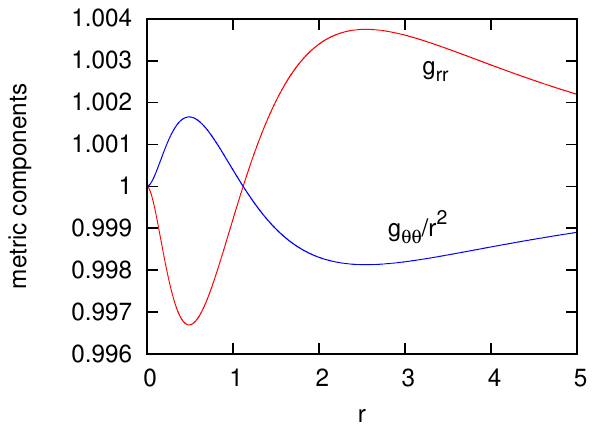}}
	\caption{Conformal metric components $g_{rr}$ and $g_{\theta\theta}/r^2$ for the trumpet 
	geometry in gamma--driver shift coordinates.}
	\label{trumpetmetric}
\end{figure}

\begin{figure}[htb] 
	{\includegraphics[scale = 1.4,viewport = 80 50 200 180]{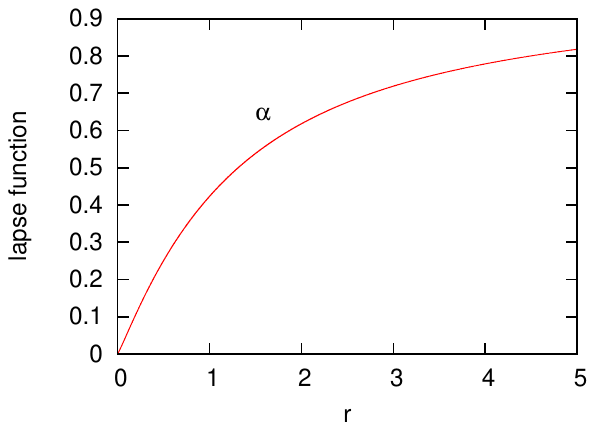}}
	\caption{Lapse function $\alpha$  for the trumpet 
	geometry in gamma--driver shift coordinates.}
	\label{trumpetlapse}
\end{figure}

\begin{figure}[htb] 
	{\includegraphics[scale = 1.4,viewport = 80 50 200 180]{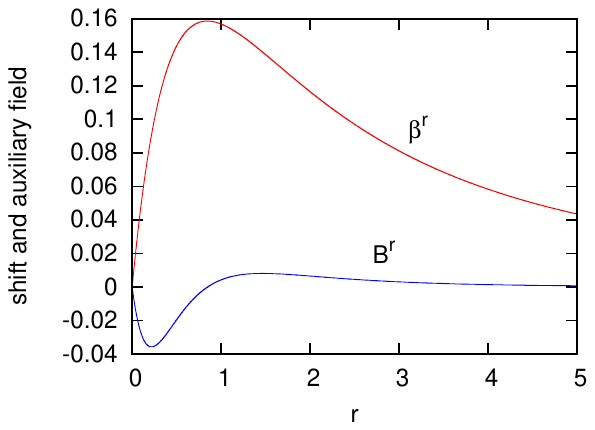}}
	\caption{Shift vector and auxiliary field components $\beta^r$ and $B^r$ 
	for the trumpet 
	geometry in gamma--driver shift coordinates.}
	\label{trumpetshift}
\end{figure}

Figure \ref{trumpetmetric} is a plot of the radial and angular components of the conformal metric 
in the stationary gamma--driver coordinates. It is difficult to tell from this figure, but both components 
have a vanishing slope at $r=0$. Figure \ref{trumpetlapse} is a plot of the lapse function and 
Fig.~\ref{trumpetshift} shows the radial components of the shift vector and the auxiliary field. 

The coordinate location of the black hole horizon in stationary gamma--driver shift coordinates is 
$r = 0.842$. (Again, as in the body of this paper, I have set $M=1$.) 
In isotropic coordinates, the horizon location is $\rho = 0.839$. 
The location where $e^{4\varphi} = 2\alpha$ and hyperbolicity breaks down is $r = 4.06$ in 
stationary gamma--driver coordinates. To three significant digits, this location has the same 
value in isotropic coordinates: $\rho = 4.06$. 

My original strategy for this research project was to use the stand--alone code to generate 
a very high resolution data set for the unperturbed trumpet. That data would then be read into 
the cartoon BSSN code and interpolated onto the grid used for evolution. This strategy does not 
work very well. One of the difficulties is that there are many places in which numerical 
errors can accumulate. In particular, we have the integration in Eq.~(\ref{ellintegral}), 
the integration of $\partial\rho/\partial\ell = \rho/R$, the solution of the elliptic 
Eq.~(\ref{ellipticeqnforr}), the transformations from isotropic to stationary gamma--driver coordinates, and finally 
the interpolations onto the evolution grid. With my particular implementation, the errors in the 
unperturbed trumpet data were large enough to keep the evolution code from showing clean 
third--order convergence.

Another way that one can generate a trumpet geometry in gamma--driver coordinates is to begin with wormhole data 
and evolve that data until it settles to a stationary state. When a wormhole evolves to a trumpet, there is an 
adjustment pulse that propagates out from the puncture boundary to infinity. 
It takes a time of about $t = 100$ for the trailing edge of this pulse to propagate beyond $r = 50$. 
To be sure that the data are not contaminated with errors from the 
outer boundary, I evolve the initial wormhole for a time of $t = 110$ on a grid with an outer boundary 
at $r_{max} = 170$. The data beyond $r = 50$ is then discarded. 

The unperturbed data produced by this second method, evolving a wormhole for $t=110$, works fairly well 
for studies with initial perturbations in modes $\chi_4^-$, $\chi_5^-$, and 
$\chi_6^-$.  In these cases, with $r_0 = 10$, the perturbations reach the puncture boundary within 
a time of $t \approx 20$ or less. A total simulation time of around $t = 25$ is sufficient to reveal
any reflections that might arise in mode $\chi_6^+$. With the outer boundary at $r_{max} = 50$, 
we can be confident that the results are not contaminated with errors from the outer boundary. 

The studies with initial perturbations in $\chi_1$, $\chi_2$, and $\chi_3$ require much longer run times, 
up to $t = 120$. 
This is because these modes move very slowly toward the puncture. For these cases, the second method 
of generating unperturbed trumpet data would require us to evolve the wormhole for
$t \approx 180$ on a grid that extends to $r_{max} \approx 300$. This is not feasible with 
my current code. Thus, in Sec.~V.D I used the first method, the method of 
Eqs.~(\ref{stationaryGDS}) through (\ref{ellipticeqnforr}), to generate unperturbed trumpet data.

The two methods of calculating unperturbed trumpet data yield similar results. In fact, 
one cannot tell the difference between the two methods by examining graphs  
such as those in Figs.~\ref{trumpetmetric}---\ref{trumpetshift}. The differences only appear 
upon more careful examination, such as convergence testing. 

Another problem with the first method  
is that, near the puncture, the data is in some sense ``too good". For much of my work I used the first 
method to generate trumpet data with 
a resolution of $\Delta r \approx 0.00069$. That data was interpolated onto an evolution grid 
with a typical resolution of around $h = 0.01$. 
With the cartoon evolution code, as with any puncture method code, the resolution is very poor near 
the puncture boundary and the truncation errors are high. On the other hand, the truncation errors 
in the data produced by the first method are relatively low near the puncture boundary. 
Because of this mismatch, the unperturbed trumpet 
data is not quite stationary when it is evolved with the evolution code. In  particular, the 
grid points near the puncture boundary evolve as they adjust to the larger truncation errors 
of the evolution code. This adjustment creates a perturbation that propagates outward through 
the computational domain. The adjustment is  
difficult to notice on a plot of the BSSN or gauge variables. However, 
the adjustment readily appears as an excitation in the characteristic fields near the puncture. 
In some cases the excitation is much larger than the perturbations we wish to study. 

I can correct for this problem in the following way. For each of the simulations with initial 
perturbations in $\chi_1$, $\chi_2$ and $\chi_3$, I ``normalize" the data by subtracting the results obtained from a 
simulation with no perturbation. This removes the adjustment pulse from the data. The normalized data is 
easier to interpret than the raw data. 

The same normalization scheme is applied to the simulations with initial perturbations in $\chi_4^-$, 
$\chi_5^-$ and $\chi_6^-$, but for a somewhat different reason. In these cases the unperturbed trumpet 
data is derived using the second method, the method of evolving a wormhole for $t=110$. However, the mode 
analysis is very sensitive to any change in the stationary trumpet data. What can be seen 
in the mode plots is that the unperturbed trumpet data  is not entirely stationary. Even 
after a run time of $t=110$ there is still a small amount of evolution taking place 
near the puncture boundary. For example, when the unperturbed trumpet data (prepared 
by the second method) is evolved for 
another $t=20$, the value of mode $\chi_6^+$ drifts from $0$ to $-2.5\times 10^{-5}$ 
near the puncture boundary. These effects are removed from the data for perturbed trumpet 
simulations by subtracting the data for an unperturbed trumpet simulation. 

To summarize, the simulations from Sec.~V that have initial perturbations in modes $\chi_4^-$, 
$\chi_5^-$ and $\chi_6^-$ use the second method for generating unperturbed trumpet data. 
The simulations that have initial perturbations in modes $\chi_1$, $\chi_2$ and $\chi_3$ 
use the first method for generating unperturbed trumpet data. In all cases, the data 
shown in Sec.~V have been normalized by subtracting the data obtained from a simulation 
of the unperturbed trumpet.

\begin{acknowledgments}
I would like to thank Bernard Kelly, Richard Price, Manuel Tiglio and James van Meter 
for helpful comments. 
This work was supported by NSF Grant No. PHY--0758116. 
\end{acknowledgments}

\bibliography{references}

\end{document}